\documentclass[10pt,journal,compsoc]{IEEEtran}
\ifCLASSOPTIONcompsoc
  \usepackage{cite}
\else
  \usepackage{cite}
\fi
\ifCLASSINFOpdf
\else
\fi
\usepackage{psfrag,amsbsy,graphics,float}
\usepackage{verbatim} 
\usepackage{url}
\usepackage{cite}

\usepackage{wrapfig}
\usepackage[ansinew]{inputenc}

\usepackage{algorithm}
\usepackage[noend]{algorithmic}
\usepackage{amsmath}
\usepackage{amssymb}
\usepackage{amsthm}
\usepackage{bbm}
\usepackage{enumerate}
\usepackage{graphicx}
\usepackage{ifthen}
\usepackage{struktex}
\usepackage{hyperref}
\usepackage{color}
\usepackage{nicefrac} 
\usepackage{multirow}
\usepackage{hhline}
\usepackage{cleveref}
\usepackage{tabularx}
\usepackage{rotating}
\usepackage{caption}
\usepackage{subcaption}
\usepackage{rotating}
\usepackage{listings}
\usepackage{xcolor}
\usepackage{xtab}
\lstdefinestyle{Python}{
    language        = Python,
    basicstyle      = \ttfamily,
    keywordstyle    = \color{blue},
    keywordstyle    = [2] \color{teal}, 
    stringstyle     = \color{green},
    commentstyle    = \color{red}\ttfamily
}

\usepackage{booktabs}
\usepackage{makecell}
\usepackage{longtable}
\usepackage[nolist]{acronym}
\usepackage[font=small,skip=5pt]{caption}

\usepackage{adjustbox}

\usepackage{soul}

\usepackage{newfloat}
\DeclareFloatingEnvironment[name={Supplementary Figure},fileext=lsf,listname={List of Supplementary Figures}]{suppfigure}

\DeclareFloatingEnvironment[name={Supplementary Tables},fileext=lsf,listname={List of Supplementary Tables}]{supptable}

\newcommand{\ie}{i.\,e. }
\newcommand{\eg}{e.\,g. }
\newcommand{\etal}{et\,al. }
\newcommand{\etalns}{et\,al.}
\newcommand{\ds}{\mbox{\textsc{Deep Spectrum }}}
\newcommand{\dsns}{\mbox{\textsc{Deep Spectrum}}}
\newcommand{\emonet}{\mbox{\textsc{EmoNet }}}

\newcommand{\emonetns}{\mbox{\textsc{EmoNet}}}

\newcommand{\emoset}{\mbox{\textsc{EmoSet }}}
\newcommand{\emosetns}{\mbox{\textsc{EmoSet}}}

\def\realnumbers{\mathbb{R}}

\begin{document}
%
\title{EmoNet: A Transfer Learning Framework for Multi-Corpus Speech Emotion Recognition}
%
%
%
%

\author{Maurice~Gerczuk,
        Shahin~Amiriparian,~\IEEEmembership{Member,~IEEE,}
        Sandra~Ottl,
        and~Bj\"orn~W.\ Schuller,~\IEEEmembership{Fellow,~IEEE}}
\IEEEtitleabstractindextext{%
\begin{abstract}
In this manuscript, the topic of multi-corpus \acf{SER} is approached from a deep transfer learning perspective. A large corpus of emotional speech data, \textbf{\emosetns}, is assembled from a number of existing \ac{SER} corpora. In total, \emoset contains \textbf{84\,181 audio recordings} from \textbf{26 \ac{SER} corpora} with a total duration of over \textbf{65 hours}. The corpus is then utilised to create a novel framework for multi-corpus speech emotion recognition, namely \textbf{\emonetns}. A combination of a deep ResNet architecture and residual adapters is transferred from the field of multi-domain visual recognition to multi-corpus \ac{SER} on \emosetns. Compared against two suitable baselines and more traditional training and transfer settings for the ResNet, the residual adapter approach enables parameter efficient training of a multi-domain \ac{SER} model on all 26 corpora. A shared model with only $3.5$ times the number of parameters of a model trained on a single database leads to increased performance for 21 of the 26 corpora in \emosetns. Measured by McNemar's test, these improvements are further significant for ten datasets at $p<0.05$ while there are just two corpora that see only significant decreases across the residual adapter transfer experiments. Finally, we make our \emonet framework publicly available for users and developers at \href{https://github.com/EIHW/EmoNet}{https://github.com/EIHW/EmoNet}. \emonet provides an extensive command line interface which is comprehensively documented and can be used in a variety of multi-corpus transfer learning settings.

\end{abstract}

}

\maketitle

\IEEEdisplaynontitleabstractindextext

%
\IEEEpeerreviewmaketitle

\IEEEraisesectionheading{\section{Introduction}\label{sec:introduction}}

%
%
%
%
\IEEEPARstart{W}ith recent advancements in the field of machine learning and the widespread availability of computationally powerful consumer devices, such as smartphones, many people are now already interacting with artificial intelligence (AI) technology on a daily basis. A prominent example are voice controlled personal assistants, such as Alexa and Siri~\cite{hoy2018alexa, lopez2017alexa}, which are becoming increasingly popular. These products are a first step towards conversational AI, which integrates a variety of research fields, such as automatic speech recognition (ASR), natural language understanding (NLU), and context modelling~\cite{ram2018conversational}. While their current functionality demonstrates their capabilities for task oriented ASR, they are still a far way off of being able to converse freely with humans~\cite{levesque2017common} and disregard other important aspects of interpersonal communication, such as the expression and understanding of emotions.

In contrast to an early view of emotions being a disruption of organised rational thought and to be controlled by an individual~\cite{young1943emotion, woodworth1940psychology, young1936motivation}, today, emotional intelligence is accepted as a central and guiding function of cognitive ability that meets standards for an intelligence~\cite{salovey1990emotional, ciarrochi2006emotional, mayer1999emotional, mayer2001emotional}. In this context, it has been argued that designing conversational AIs with both the ability to comprehend and to express emotions will improve on the quality and effectiveness of human machine interaction~\cite{becker2007emotions, ball2000emotion}. Emotion recognition capabilities can furthermore serve a purpose for the integration of machine learning and AI technologies into health-care where an affective computing approach can support patient in-home monitoring~\cite{mano2016exploiting}, help with early detection of psychiatric diseases~\cite{tacconi2008activity} or be applied to support diagnosis and treatment in military healthcare~\cite{tokuno2011usage}. In this regard, \ac{SER} is furthermore highly related to automatic speech based detection of clinical depression~\cite{cummins2011investigation, low2010detection, cummins2015review, cummins2013diagnosis}, where emotional content can further improve the performance of recognition approaches~\cite{stasak2016investigation}. Here, monitoring of patients using only audio signals could be seen as less intrusive than video or physiological based methods. 

While \ac{SER} can serve a wide range of purposes from an application stand-point, there are a couple of fundamental characteristics of the field that make it a hard task up to this day~\cite{schuller2018speech}.
Collecting and annotating sufficiently large amounts of data that is  suitable to the target application is time consuming~\cite{schuller2018speech}. 
Emotional speech recordings can for example be obtained in laboratory setting by recording professional actors or in the wild~\cite{avila2018feature}, \eg, from movie scenes or online videos~\cite{koolagudi2012emotion}. Moreover, after suitable material has been collected, annotation is not straightforward and has to consider an additional modelling step in the different ways human emotions can be classified and interpreted~\cite{schuller2018speech}. Here, choosing between a categorical and dimensional approach and defining how to represent the temporal dynamics of emotion are two examples of important steps that have to be taken to arrive at a fitting an emotion model~\cite{koolagudi2012emotion,mollahosseini2017affectnet}. The actual process of annotation then  has to find ways to deal with the inherent subjectivity of human emotion, \eg, by employing multiple annotators, measuring inter-rater agreement and removing data samples containing emotion portrayals that were too ambiguous~\cite{schuller2018speech}. For these reasons, unlike in many fields that have seen leaps in the state-of-the-art due to the paradigm of deep learning, truly large-scale -- \ie, more than one million samples -- \ac{SER} corpora do not exist up to this day. Rather, \ac{SER} research brought forth a large number of small-scale corpora which showcase a high degree of diversity in their nature.  

The sparsity of large \ac{SER} datasets on the one hand, and the availability of a large number of smaller datasets on the other, are key motivating factors for our work which applies state-of-the-art paradigms of both transfer and deep learning for the task of multi-corpus \ac{SER}. For this, we develop a multi-corpus framework called \emonet based based on residual adapters~\cite{rebuffi2018efficient,rebuffi2017learning}. We evaluate this framework on a large collection of existing \ac{SER} corpora assembled and organised as \emoset. Further, all experiments are compared to suitable baseline \ac{SER} systems. Finally, we release the source code of our framework together with a comprehensive command line interface on GitHub\footnote{\href{https://github.com/EIHW/EmoNet}{https://github.com/EIHW/EmoNet}}, to be freely used by any researcher interested in multi-domain audio analysis.

\section{Related Works} \label{chap:background}
Three general research areas are of interest for the topic of this manuscript. First, the field of \ac{SER} deals with the general problem of recognising human emotions from speech recordings, which is the task that should be solved on a multi-corpus basis by the presented approaches. Second, the field of deep learning where large amounts of data are used to train neural network architectures that are able to learn high-level feature representations from raw input data, gives the direction for the choice of machine learning models. Here, the \ds feature extraction system is of special importance, as it will serve as one of the baselines against which the transfer learning experiments are compared. Finally, the field of domain adaptation and multi-domain training investigates ways of efficiently transferring knowledge between domains which is important for the multi-corpus \ac{SER} problem. 

\begin{figure*}
\begin{center}
\includegraphics[width=.8\textwidth]{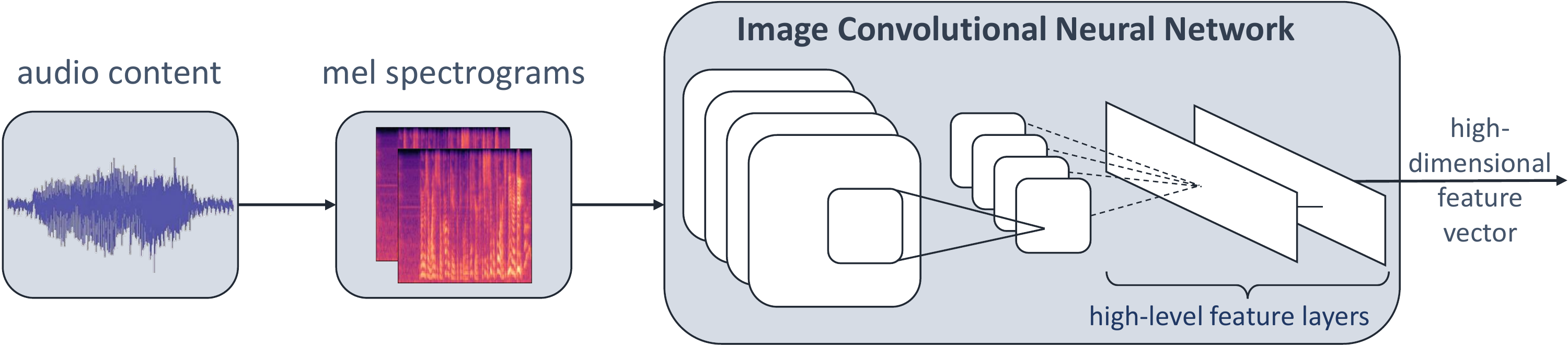}
\end{center}
\caption{Overview of the \ds feature extraction system. Audio segments are first converted to a suitable image representation in the form of mel-spectrogram plots. Afterwards, they are forwarded through an ImageNet pre-trained deep \ac{CNN} model. The activations on a specific layer can then be used as high-level feature representations by downstream classifiers, such as linear \acp{SVM}. This figure is adapted from~\cite{Amiriparian18-BND}.}
\label{fig:deepspectrum}
\end{figure*}

\subsection{Speech Emotion Recognition}
\label{sec:ser}
\ac{SER} describes the task of automatically detecting emotional subtext from spoken language and is a research field that has increasingly been of interest for more than two decades~\cite{schuller1988emotion}. Traditionally, to develop a computational emotion recognition model from speech data, the following steps should be considered. Initially, unlike for many other current machine learning tasks, \eg, visual object recognition or \ac{ASR}, no single definitive method of representing human emotion is applicable to every scenario
~\cite{schuller2018speech}. Furthermore, a decision has to be made about the temporal granularity of annotating and detecting emotions from continuous speech signals~\cite{schuller2018speech}. For the first aspect, two main approaches are used in the literature~\cite{gunes2013categorical}. In the categorical approach, emotions can be grouped into discrete classes~\cite{ekman1999basic}, mostly with respect to the Ekman `Big Six'~\cite{ekman1984expression, ekman1994nature}, including 
anger, disgust, fear, happiness, sadness and surprise. On the other hand, emotions can  be analysed from a continuous and dimensional perspective. Here, an emotion is annotated on multiple axes each describing a different aspect. Most often used dimensions are \emph{arousal}, \emph{valence} and \emph{dominance}. Arousal describes the intensity or activation of an emotion, while valence represents the intrinsic positivity or negativity of a feeling~\cite{frijda1986emotions}. Finally, emotions can also be placed on a continuous dominance-submissiveness scale which, for example, allows discrimination between anger and fear (both low valence high arousal emotions)~\cite{mehrabian1980basic}. Translation between categories and dimensions is possible, \eg, using a mapping, such as the circumplex model of affect~\cite{russell1980circumplex}. 

For the actual automatic computational recognition of emotions from audio speech signals, many traditional approaches rely on the extraction of either brute-forced or handcrafted sets of features computed from frame-level \acp{LLD}, such as energy or spectral information, by applying a range of statistical functionals, such as mean, standard deviation, and extreme values over a defined longer segment, \eg, a speaker turn or utterance~\cite{el2011survey, anagnostopoulos2015features}. More recently, the field is  being influenced by the trend of deep learning and methods of directly learning models for emotion recognition from raw speech data are being investigated~\cite{schuller2018speech,ottl2020group}. These works are touched upon in~\Cref{sec:deep-learning}.

In the computational and machine learning sphere, previous works have shown the suitability of cross-corpus training for \ac{SER}~\cite{schuller2011selecting, schuller2011using}. Aspects that were researched include how to effectively select data from multiple corpora~\cite{schuller2011selecting, schuller2011using} or the effects of adding unlabelled data for training~\cite{zhang2011unsupervised}. Schuller \etalns~\cite{schuller2010cross} have investigated various strategies to cope with inter corpus variance by evaluating multiple normalisation strategies, such as per-speaker and per-corpus feature normalisation, for cross-corpus testing on six databases while Kaya \etalns~\cite{kaya2018efficient} applied cascaded normalisation on standard acoustic feature sets. Many of the above approaches utilise hand-crafted or brute-forced feature sets computed from \acp{LLD} of the audio content. The approach taken in this manuscript differs from the previous research into cross-corpus \ac{SER}, in that the focus here lies on harnessing the power of deep learning when applied to raw input features.

\subsection{Deep Learning Based SER}
\label{sec:deep-learning}
For computer audition,  deep learning has made an impact -- especially in the \ac{ASR} domain where large datasets of recorded speech and corresponding transcriptions could be utilised to train accurate models~\cite{hannun2014deep,amodei2016deep}. 

In comparison to image recognition where training samples can be collected in a large-scale fashion from the internet and annotation is relatively straight forward, many areas of audio recognition do not have this advantage. In an attempt to improve the situation for audio recognition, AudioSet~\cite{gemmeke2017audio} which is a large ontology of collected YouTube clips 
can serve as basis for training deep, general purpose network architectures. These networks can later be used as feature extractors and transfer models for various recognition tasks. Hershey \etal investigated the viability of adapting standard \ac{CNN} architectures as used for ImageNet classification, such as VGG~\cite{simonyan2014very}, Inception~\cite{szegedy2016rethinking}, and ResNet~\cite{he2016deep, szegedy2017inception}, for large-scale audio recognition~\cite{hershey2017cnn}. Unlike in image recognition, where the raw pixels of (resized) images can serve as input to the \acp{DNN}, a suitable data representation for audio content has to be chosen manually. Additionally, when the machine learning models cannot handle variable size inputs, \eg, \acp{CNN} with fully connected layers, audio has to be chunked into segments of fixed duration. Tzirakis \etalns~\cite{tzirakis2018end} took a slightly different approach by introducing a \ac{CRNN} architecture that transforms the audio signal to a suitable representation using one-dimensional convolutional layers. This time-continuous representation is then processed by a \ac{RNN} which makes predictions for the learning task at hand. The whole model can be trained in an end-to-end fashion from raw audio signals~\cite{amiriparian2019deep,Amiriparian18-ROE}. 

Mel-spectrograms are often used as input features for the \acp{DNN}~\cite{neumann2017attentive,mirsamadi2017automatic,girdhar2017attentional,zhang2018attention}. As not all time-frequency regions of speech samples contain high emotional saliency, methods have been investigated that learn to focus on the most important parts of a given input. \acp{RNN}, for example, have the inherent capability to deal with sequences of variable lengths. When combined with a self attention mechanism, emotionally informative time-segments of an input can be highlighted~\cite{lin2020efficient}. Mirsamadi \etal used an attention \ac{RNN} for \ac{SER} on \ac{IEMOCAP} while Gorrostieta \etal applied a similar model with low-level spectral features as input for the ComParE self-asessed affect~\cite{schuller2018interspeech} sub-challenge~\cite{gorrostieta2018attention}. More recently, combining \ac{CNN} feature extractors with attention based \acp{RNN} has been shown to be a highly competitive approach to \ac{SER}~\cite{latif2020deep, fujioka2020meta}. Taking inspiration from the combination of \acp{FCN} and two-dimensional attention pooling for visual online object tracking~\cite{chu2017online}, Neumann and Vu evaluate an attentive \ac{CNN} for \ac{SER} on different input features and find log Mel-spectrograms to work best. Their work  highlights another advantage an attention pooling method has compared to a more traditional \ac{CNN} architecture with fully connected layers: Sequences of variable lengths can be handled by the same model without having to adapt parts of the architecture. This is especially important for \ac{SER} where utterances often differ in duration, and analysing only a short segment of a long utterance often leads to worse results. While the works outlined above demonstrate the efficacy of deep learning based methods for \ac{SER} on various databases, hand-crafted features still play an important role in the field~\cite{wagner2018deep}.

\begin{figure*}[tph]
\begin{center}
\includegraphics[width=0.9\textwidth]{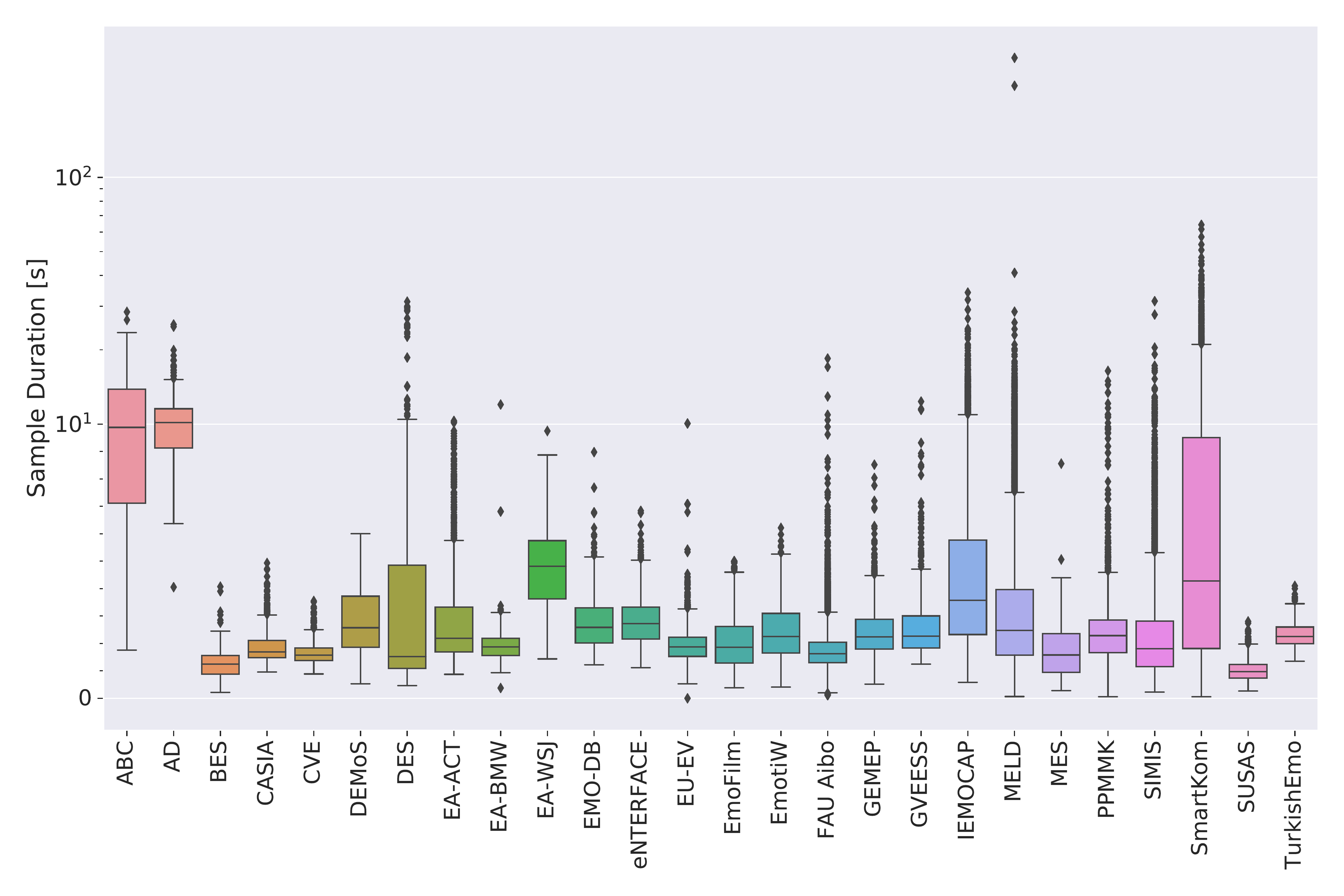}
\end{center}
\caption{Boxplot of sample durations for each \emoset corpus. Boxes show the inner-quartile range (IQR) and the whiskers extend to a maximum of $1.5\times IQR$ measured from the lower and higher quartiles. Black dots are considered as outliers. For readability, the scale of the y-axis is logarithmic from $10^{1}$ upwards.}
\label{fig:duration-box}
\end{figure*}

\subsection{Deep Spectrum Feature Extraction}
\label{sec:ds-background}
First introduced for the task of snore sound classification~\cite{Amiriparian17-SSC}, \dsns\footnote{\href{https://github.com/DeepSpectrum}{https://github.com/DeepSpectrum}} is a deep feature extraction toolkit for audio-based on image classification CNNs. In the system, knowledge obtained by \acp{CNN} for the task of large-scale object recognition on the ImageNet~\cite{deng2009imagenet} database is transferred to the audio domain by applying the learnt feature extractors to spectrogram representations. Specifically, an audio sample is first converted to a suitable 2-dimensional time-frequency format -- most often a (Mel-)spectrogram or a chromagram -- by means of Fourier transformation and application of various filter and scaling operations. Afterwards, an RGB image representation conforming with the input specifications of an ImageNet pre-trained \ac{CNN}, \eg, AlexNet~\cite{krizhevsky2012imagenet}, VGG16~\cite{simonyan2014very}, or ResNets~\cite{he2016deep}, is created by plotting and resizing the spectrogram, mapping power to a certain pre-defined colour scale. This image is then forwarded through the \ac{CNN} and the learnt filter operations are applied to the spectrogram. A specific layer of the extractor network can finally be chosen to serve as a deep feature descriptor, \ie, the activations of this layer are flattened into a (large) feature vector which can then be used as input for various machine learning models. The whole process is  illustrated in \Cref{fig:deepspectrum}.

The features extracted by the \ds system have been shown to provide state-of-the-art results for a range of audio analysis tasks in preceding research~\cite{Cummins17-AID,cummins2018multimodal,Amiriparian17-SAU,amiriparian2019mediaeval,Amiriparian19-AYP, Amiriparian18-BND,ottl2020group, amiriparian2020towards}. The results imply that convolutional filters trained on natural image classification can extract useful features for audio analysis tasks if applied to spectral representations of the content. This motivates using \ds as one of the baseline systems to evaluate on \emoset. 

\subsection{Domain Adaptation and Multi-Domain Learning}
\label{sec:domain-adapt}
As part of the larger field of transfer learning, domain adaptation aims to transfer knowledge learnt on a fully labelled source domain to a target domain where labels are either unavailable or sparse by mitigating the negative impact of domain shift~\cite{tzeng2017adversarial}. Specifically, it deals with the problem that machine learning models trained on large-scale datasets are sensitive to dataset bias~\cite{quinonero2008covariate, gretton2009dataset, sugiyama2017dataset}. 
As deep learning approaches can be seen as the state-of-the-art in many machine learning research areas, current domain adaptation research focuses on learning to map deep feature representations into a shared space~\cite{tzeng2017adversarial}. One of the most prominent examples can be found in the work of Ganin \etalns~\cite{ganin2014unsupervised}, where a shared feature extractor base serves as input for both a source domain label predictor and a domain discriminator. Similarly, Tzeng \etalns\cite{tzeng2017adversarial} apply the concept of adversarial learning to unsupervised domain adaptation by first pre-training a source domain feature encoder in a supervised manner. In the context of \ac{SER}, domain adaptation has been investigated for the purposes of increasing performance on domains where labelled data is sparse or performance it hindered by domain shifts. Ramakrishnan \etalns~\cite{abdelwahab2015supervised} used supervised domain adaptation to adapt \ac{SVM} classifiers from two different source domains (\ac{IEMOCAP} and SEMAINE) to 
RECOLA. Hassan \etalns~\cite{hassan2013acoustic} further improved on the best challenge submissions for \ac{AIBO} by mitigating the domain shift caused by the two different recording sites of samples in the database via importance weighting for \acp{SVM}.

While most of the work done for domain adaptation focuses on transferring models and feature representations from a single source domain to a new target domain, the setting of multi-corpus \ac{SER} as found in this manuscript is different. Due to unavailability of truly large-scale databases in the field, an approach that is able to learn from multiple domains at once and in the process increases performance for individual tasks is more desirable. Following work into the area of domain adaptation training, the model of residual adapters was first introduced in~\cite{rebuffi2018efficient} for the task of multi-domain learning in the visual domain.
The approach tries to find a middle ground between using large pre-trained networks as feature extractors and finetuning the networks on a new task. While using pre-trained networks as feature extractors might have performance drawbacks, finetuning the whole network is very parameter inefficient and can easily lead to catastrophic forgetting of the model's pre-training. Instead, in their approach, not a single universal multi-domain model is trained, but families of networks that are highly related to one another, sharing a large portion of their parameters while still containing task specialised modules. In their work on the Visual Decathlon~\cite{rebuffi2017learning} challenge, the authors experimented with different placements of and configurations for the adapter modules. They found that for the best performance, adapters should be used throughout the whole deep network in a parallel configuration. In general, this method reaches or even surpasses traditional finetuning strategies while only requiring about $10\,\%$ of task specific parameters. Due to the highly promising results on different domains of visual recognition and the capabilities of \acp{CNN} for various speech recognition tasks -- as touched upon in~\Cref{sec:ds-background} and~\Cref{sec:deep-learning} -- the model is adapted for multi-corpus \ac{SER} for this manuscript.

\begin{table*}
\footnotesize
\centering
\caption{Statistics of the \emoset databases in terms of number of speakers (\emph{Sp.}) and classes (\emph{C.}), mean sample duration in seconds (\emph{mean\_dur}), total duration (\emph{total\_d}) and spoken language(s).} 
\label{tab:emoset-stats}
\resizebox{\textwidth}{!}{
\begin{tabular}[t]{lrrrlrr}

\toprule
                    \textbf{name} &  \textbf{\#} & \textbf{\# C.} &\textbf{\# Sp.} & \textbf{Language} &  
                    \textbf{mean\_dur [s]}&  \textbf{ total\_d [h]} 
                    \\
\midrule
\textbf{\ac{ABC}}~\cite{schuller2007audiovisual} &  405 & 5 & 8 & German 
 & $10.6 \pm 4.7$ & 1.19
 \\ 
 
 \textbf{\ac{AD}} (cf. \Cref{ssec:ad}) & 660 & 2 & 9 & German 
 & $10.5 \pm 2.2$ & 1.93 
 \\ 
 \textbf{\ac{BES}}\cite{nwe2003speech} &  414 & 6 & 6 & Burmese 
 & $1.3 \pm 0.5$ & 0.15 
 \\ 
 \textbf{\acs{CASIA}}~\cite{CASIA} &  1200 & 6 & 4 & Mandarin 
 & $1.9 \pm 0.6$ & 0.64 
 \\ 
 \textbf{\ac{CVE}}~\cite{liu2012recognizing} &  874 & 7 & 4 & Mandarin 
 & $1.7 \pm 0.4$ & 0.40 
 \\ 
 \textbf{\ac{DEMoS}}~\cite{parada2019demos} &  9\,365 & 7 & 68 & Italian
 & $2.9 \pm 1.3$ & 7.40 
 \\ 
 \textbf{\ac{DES}}~\cite{engberg1997design} &  419 & 5 & 4 & Dutch 
 & $4.0 \pm 5.7$ & 0.47 
 \\ 
 \textbf{EA-ACT}~\cite{schuller2006automatische} &  2\,280 & 7 & 39 & \makecell[tl]{English\\French\\German}
 & $2.8 \pm 1.7$ & 1.80 
 \\ 
\textbf{EA-BMW}~\cite{schuller2006automatische} &  1\,424 & 3 & 10 & German 
 & $1.9 \pm 0.5$ & 0.76 
 \\ 
 \textbf{EA-WSJ}~\cite{schuller2006automatische} &  520 & 2 & 10 & English 
 & $4.7 \pm 1.5 $ & 0.69 
 \\ 
 \textbf{\ac{EMO-DB}}~\cite{burkhardt2005database} &  494 & 7 & 10 & German 
 & $2.8 \pm 1.0$ & 0.38 
 \\ 
 \textbf{\ac{EmotiW}}~\cite{dhall2014emotion} &  961 & 7 & - & -  
 & $2.4 \pm 1.0$ & 0.65 
 \\ 
\textbf{\ac{eNTERFACE}}~\cite{martin2006enterface} &  1\,277 & 6 & 42 & English 
 & $2.8 \pm 0.9$ & 1.00 
 \\ 
 \textbf{\ac{EU-EV}}~\cite{Lassalle2019} & 3\,148 & 21 & 54 & \makecell[tl]{English\\Hebrew\\Swedish} 
 & $1.9 \pm 0.8$ & 1.70 
 \\  
\textbf{EmoFilm}~\cite{parada2018categorical} &  1\,115 & 5 & 207 & \makecell[tl]{English\\Italian\\Spanish}
 & $2.1 \pm 1.0$ & 0.64 
 \\  
 \textbf{\ac{AIBO}}~\cite{batliner2008releasing} &  17\,074 & 5 & 51 & German 
 & $1.7 \pm 0.7$ & 8.25 
 \\ 
 \textbf{\ac{GEMEP}}~\cite{banziger2010introducing, banziger2012introducing} &  1\,260 & 17 & 10 & French 
 & $2.4 \pm 1.0$ & 0.85 
 \\ 
 \textbf{\ac{GVEESS} (Full Set)}~\cite{banse1996acoustic} &  1\,344 & 13 & 12 & Made-up 
 & $2.6 \pm 1.1$ & 0.96 
 \\ 
 \textbf{\ac{IEMOCAP} (4 classes)}~\cite{busso2008iemocap} &  5\,531 & 4 & 10 & English
 & $4.6 \pm 3.2$ & 7.00 
 \\ 
 \textbf{\ac{MELD}}~\cite{poria2018meld} & 13\,707 & 7 & 6+ & English 
 & $3.2 \pm 4.0$ & 12.10 
 \\ 
 \textbf{\ac{MES}}~\cite{nwe2003speech} &  360 & 6 & 6 & Mandarin 
 & $1.8 \pm 1.0$ & 0.18 
 \\ 
 \textbf{PPMMK} (cf. \Cref{ssec:ppmmk}) &  3\,154 & 4 & 36 & German 
 & $2.5 \pm 1.3$ & 2.16 
 \\ 
 \textbf{\ac{SIMIS}}~\cite{schuller2018speech, schuller2020speech} &  9\,299 & 5 & 10 & German 
 & $2.3 \pm 1.7$ & 5.80 
 \\ 
 \textbf{\ac{SmartKom}}~\cite{schiel2002smartkom} &  3\,823 & 7 & 79 & German
 & $6.7 \pm 7.0$ & 7.10 
 \\ 
 \textbf{\ac{SUSAS}}~\cite{hansen1997getting} &  3\,593 & 4 & 7 & English 
 & $1.0 \pm 0.4$ & 1.00 
 \\ 
 \textbf{TurkishEmo} (cf. \Cref{ssec:turk-emo}) &  484 & 4 & 11 & Turkish 
 & $2.3 \pm 0.5$ & 0.31 
 \\ 
\bottomrule
\end{tabular}
}
\end{table*}

\section{EmoSet - Collection of Speech Emotion Recognition Corpora}
\label{sec:emoset}
For this manuscript, a large collection of 26 emotional speech corpora has been assembled and organised. This collection, from hereon \emosetns, contains published \ac{SER} databases that have been used for research and experiments in the field. Additionally, some unpublished speech databases of the Chair of Embedded Intelligence for Healthcare and Wellbeing (University of Augsburg) are used to further augment the training data in order to improve the generalisation capabilities and minimise the overfitting problems of deep learning models. The individual datasets  had to be structured for training. Here, speaker-independent training, development, and test partitions had to be constructed manually, in the case they did not already exist.
An overview of each corpus can be found in~\Cref{tab:emoset-stats}. For published corpora, descriptions can be found in the referenced papers whereas unpublished databases are described below. These corpora all include categorical emotion labels. Overall, there are $84\,161$ samples with a total duration of $65.6$ hours and the mean duration of all audio recording is $2.81$ seconds.

\subsection{\acf{AD}}
\label{ssec:ad}
\acf{AD} is a corpus of angry and neutral speech recorded in the setting of phone calls. The corpus is split over 9 calls, each containing speech segments for both classes. In total, there are 660 samples with an average duration of $10.5$ seconds. Two calls are separated from the training data, one for the development partition and one for the held-out test set.

\subsection{PPMMK-EMO}
\label{ssec:ppmmk}
PPMMK-EMO is a database of German emotional speech recorded at the University of Passau covering the four basic classes angry, happy, neutral, and sad. It has a total of 3\,154 samples averaging 2.5 seconds in length recorded from 36 speakers. For the test set, 8 speakers' recordings are set aside, while $20\,\%$ of the training data is randomly sampled to form the development partition.
\subsection{Turkish Emotion}
\label{ssec:turk-emo}
This Turkish \ac{SER} corpus contains 484 samples of emotional speech recorded from 11 subjects (7 female and 4 male) covering the basic classes anger, joy, neutrality, and sadness. The samples have an average duration of 2.3 seconds. Two female and two male speakers' samples are used set aside for the development and test partition.

\subsection{Exploratory Data Analysis}
While a number of statistics for each \emoset corpus can be found in~\Cref{tab:emoset-stats}, further data exploration can lead to higher level insights into the nature and composition of \emosetns.
How much data is provided by each individual database, both in terms of number of samples and total duration of audio content, is highly variable.
Here, a few datasets stand out by including a comparatively large amount of data. \ac{DEMoS}, \ac{AIBO}, \ac{IEMOCAP}, \ac{MELD}, and \ac{SIMIS} all contain more than $5\,000$ samples. When looking at these corpora from the perspective of their total duration, the variability of sample lengths shows its effect. While \ac{SIMIS} includes more samples than \ac{IEMOCAP} (almost twice as many), they are a lot shorter on average, leading to a smaller combined corpus duration. A similar picture is found for \ac{AIBO} and \ac{MELD}, where the latter contains very long samples while the former's speech recordings are shorter on average.

\begin{figure}[tph]
\begin{center}
\includegraphics[width=1\columnwidth]{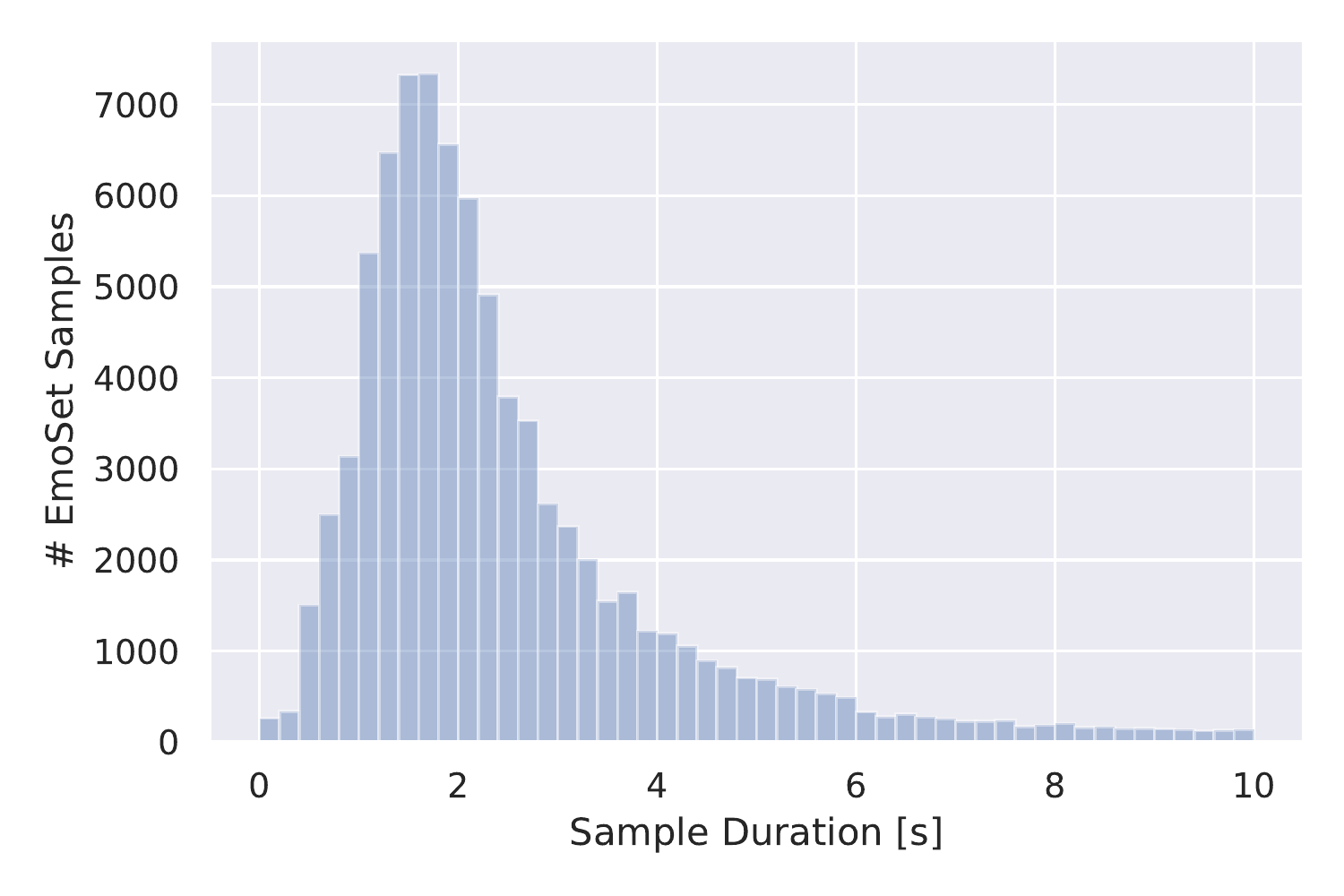}
\end{center}
\caption{Histogram of sample durations in \emosetns. Most of the samples are between 1 to $5$ seconds in length.}
\label{fig:duration-histogram}
\end{figure}

In this context, it is further of interest to take a look at the distribution of the length of audio recordings in the whole \emoset database. ~\Cref{fig:duration-histogram} shows a histogram of all sample durations from 0 to 10 seconds -- there are  a few much longer samples, but they are excluded for keeping the histogram readable. From the plot, it can be seen that most of the samples are between 1 and 5 seconds long, \ie, when choosing a window size for emotion classification later on, 5 seconds should be enough to capture a sufficiently large portion of the the input samples in their entirety.

Observations about the samples in each dataset regarding their duration can be made on a more granular level from~\Cref{fig:duration-box} which shows a boxplot for each of the 26 corpora. Here, most datasets exhibit relatively narrow duration ranges in the order of a few seconds. However, the variability of sample durations depends on the recording setting and nature of the databases, with acted and scripted emotion portrayals having a very narrow inner-quartile range while natural or induced emotional speech samples vary more heavily in their length. Thus, it is interesting to look at the databases which stick out from the pack. First of all, \ac{ABC} has samples that are 10 seconds long on average but shows a large inner-quartile range and minimum durations of under 2 seconds. The variability here can be explained by the nature of the dataset. \ac{ABC} contains unscripted induced emotional speech recorded in the simulated setting of an airplane. Emotional speech segments were annotated and extracted by experts after the recording, leading to no bounds on sample lengths. Similarly for \ac{AD}, the speaker turns of angry and neutral speech during phone conversations are generally longer than for other datasets at 10 seconds. This seems reasonable  considering that in phone calls, reliance on the actual content of speech is greater to convey one's intentions than in face to face conversations, where additional information is transported through gestures and facial expressions. \ac{DES} can be seen as a further exception here. While the corpus contains only acted and scripted emotion portrayals, the fact that there are different types of spoken content -- short utterances, sentences and fluent passages of text -- leads to a very large range of sample durations. The popular benchmark dataset \ac{IEMOCAP} also shows high variability in this context for another reason. In the database, there are both scripted and improvised sessions of conversations between professional actors. While sample durations for scripted portrayals can be kept quite consistent in length, speaker turns in natural conversations do not conform to any restrictions. \ac{MELD} which consist of scenes from the popular TV series `Friends' sitcks out by containing very long samples -- some well over a minute long while finally \ac{SmartKom} shows the highest variability in duration from only a small fraction of a second to minute long recordings captured from the test subjects during interaction with a simulated human-machine interface.

For the implemented deep learning methods, sample duration variability can be a problem. The \ds model needs a fixed length inputs for extracting feature representations from its fully connected layer, resizing the generated spectrograms might lead to a loss in information and make it harder to learn emotionally salient information across databases. For this reason, the approach works with fixed windows of 1 second. While the 2D attention model used in the ResNet based approach enables variable size input spectrograms, duration differences might lead to problems here as well -- and similarly for the fixed window in \dsns. A very long sample which has been annotated with only one emotional category as a whole might not exhibit this emotion uniformly across its entirety. This can introduce noise into the learning process which might limit the learning capabilities of the investigated models.

\normalsize

\section{Baselines}
\label{sec:baselines}
Two baseline systems using unadapted \ds and eGeMAPS~\cite{eyben2015geneva} features for emotion recognition are included to compare the performance of transfer learning models trained on \emosetns. Additionally, for the experiments with a residual adapter \ac{CNN} model, a ResNet is trained from scratch on each of the tasks.
\subsection{eGeMAPs}
\label{ssec:egemaps}
The first baseline system uses the \ac{eGeMAPS} extracted with the help of openSMILE. Afterwards, zero mean and unit standardisation is applied to the feature sets. Here, the normalisation parameters are learnt
on the training partitions and then fixed to later be applied to the validation and test splits. A linear \ac{SVM} is then trained for the individual classification tasks. Its complexity parameter is optimised on the validation partitions of each \emoset corpus on a logarithmic scale from $1$ to $10^{-9}$. With optimised parameters, the \ac{SVM} is fit again on the combined training and development splits and then evaluated on the test partition of each dataset.
\subsection{Deep Spectrum}
\label{ssec:ds-baseline}
The second baseline is constructed from the \ds system. Features are extracted from Mel-spectrogram plots (128 Mel-filters) using VGG16 pre-trained on ImageNet as feature extractor. The plots use the magma colourmap to convert the 2D spectrogram representation to an RGB colour-coded image (see~\Cref{sec:features} for more details). No windowing of the input audio samples is applied, \ie, Mel-spectrogram plots are created from the full duration of each sample. Afterwards, this image is resized to $224\times224$ to conform with the training images of the ImageNet database. As with the eGeMAPs features, a linear \ac{SVM} classifier is used with the \ds features and its parameters are optimised as in \Cref{ssec:egemaps}. 

\subsection{ResNet from Scratch}
\label{ssec:baselines-resnet}
The performance of transfer experiments with the residual adapters model will be compared against an identically structured ResNet which has been trained from scratch on each of \emosetns's corpora, individually. The description of the model's architecture can therefore be found in~\Cref{ssec:parallel-res-adapt}. The model is optimised in batches of $64$ by \ac{SGD} with a momentum of $0.9$ and a learning rate decay of $10^{-6}$ to the weights of all layers. Further, class weighted cross entropy is used as loss to counteract the class imbalance in many datasets, and a l2 regularisation loss term is added with factor $10^{-6}$. As \ac{BN} is used throughout the network, the training starts with a large initial learning rate of $0.1$ which is exponentially decayed by factor 10 when no improvement in validation \ac{UAR} occurs for 50 epochs. This learning rate decay step happens twice and afterwards the model is trained until no \ac{UAR} improvement is seen for another 50 epochs.

\section{Deep Learning Architectures}
For the transfer learning experiments, two deep learning architectures are considered. An ImageNet pre-trained VGG16 network as used in the \ds system is finetuned on \emoset and the approach of parallel residual adapters as described in~\Cref{sec:domain-adapt} and previously used for multi-domain visual recognition is adapted and evaluated on the multi-corpus \ac{SER} problem posed by \emosetns. 
\label{sec:dl-arch}

\subsection{Preprocessing}
\label{sec:features}
For the transfer learning experiments, the input audio content of the \ac{SER} data is transformed and pre-processed. Mel-Spectrograms are derived from the spectrogram audio representation. Typically, for speech recognition tasks a window size of around $25\,$ms is chosen~\cite{hershey2017cnn, schuller2018speech} for the \ac{FFT}. For efficient computation, it is best to chose a power of two (in number of samples) as window size, \ie, an \ac{FFT} window with $512$ samples leads to a close $32\,$ms for the \emoset databases that have a fixed (resampled) sampling rate of $16\,000\,$Hz. The windows are further shifted with a hop size of $256$ samples along the input audio signal, leading to a window overlap of $50\,\%$. 

Mel-Spectrograms apply dimensionality reduction to the log-magnitude spectrum with the help of a mel-filter. In the transfer learning experiments with residual adapter models, 64 mel-filters are chosen, as it has been found that a larger number of filters often leads to worse results when used as input for a \ac{CNN}~\cite{zhao2019exploring, zhao2018deep}. The spectrograms do not have to be resized, as the usage of a two  dimensional attention module, which is attached to the fully convolutional feature extractor base, enables handling of variable size input. Nevertheless, an upper bound of $5\,s$ is set on the length of the audio segments from which spectrograms are extracted. This is well above the average duration of an \emoset sample, but some datasets contain samples which are longer than $30\,s$. In the case of such a sample, a random $5\,s$ chunk is chosen for feature extraction. As the extraction is done on the fly and on a GPU, this serves as a form of augmentation and should help against overfitting. Shorter samples are converted to spectrograms as they are, and zero-padding is only performed on a per-batch basis to the maximum sample length within this batch.

\begin{figure*}
    \centering
    \includegraphics[width=0.9\textwidth]{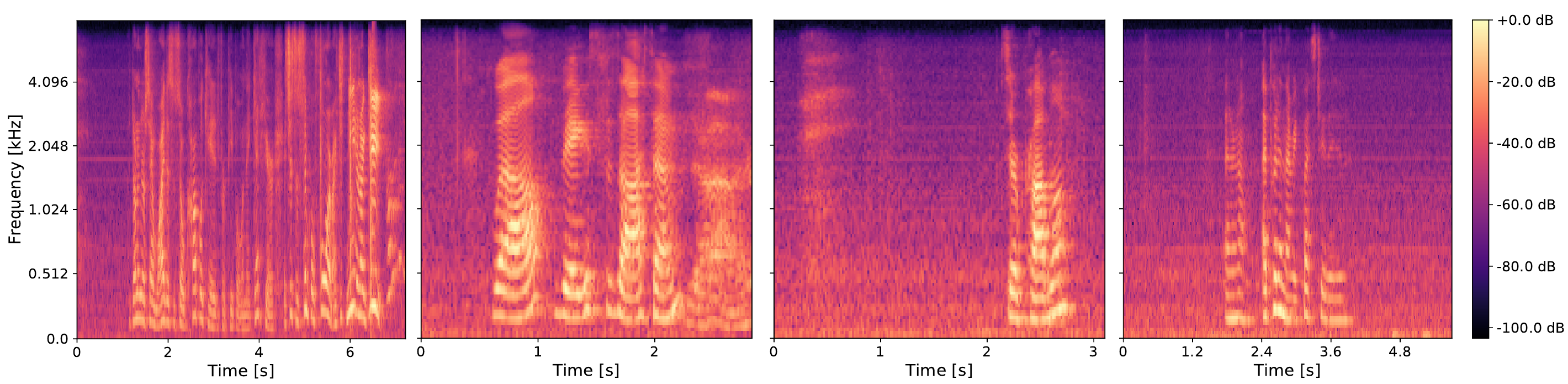}
    \caption{Sample Mel spectrogram images created from speech recordings of \ac{IEMOCAP} for each of its four base emotion categories. From left to right: angry, happy, neutral, and sad.
    }
    \label{fig:specs}
\end{figure*}

\subsection{Parallel Residual Adapters}
\label{ssec:parallel-res-adapt}
The second Deep Learning model investigated for this manuscript is built upon the concept of residual adapters, as described in~\Cref{sec:domain-adapt}.
A residual \ac{CNN} based on the popular Wide ResNet~\cite{zagoruyko2016wide} model with the parallel adapter configuration as used in~\cite{rebuffi2018efficient} is trained on \emosetns.

\subsubsection{Architecture}
\label{sssec:resnet-arch}
\begin{figure*}[tp]
\begin{center}
\includegraphics[width=0.8\textwidth]{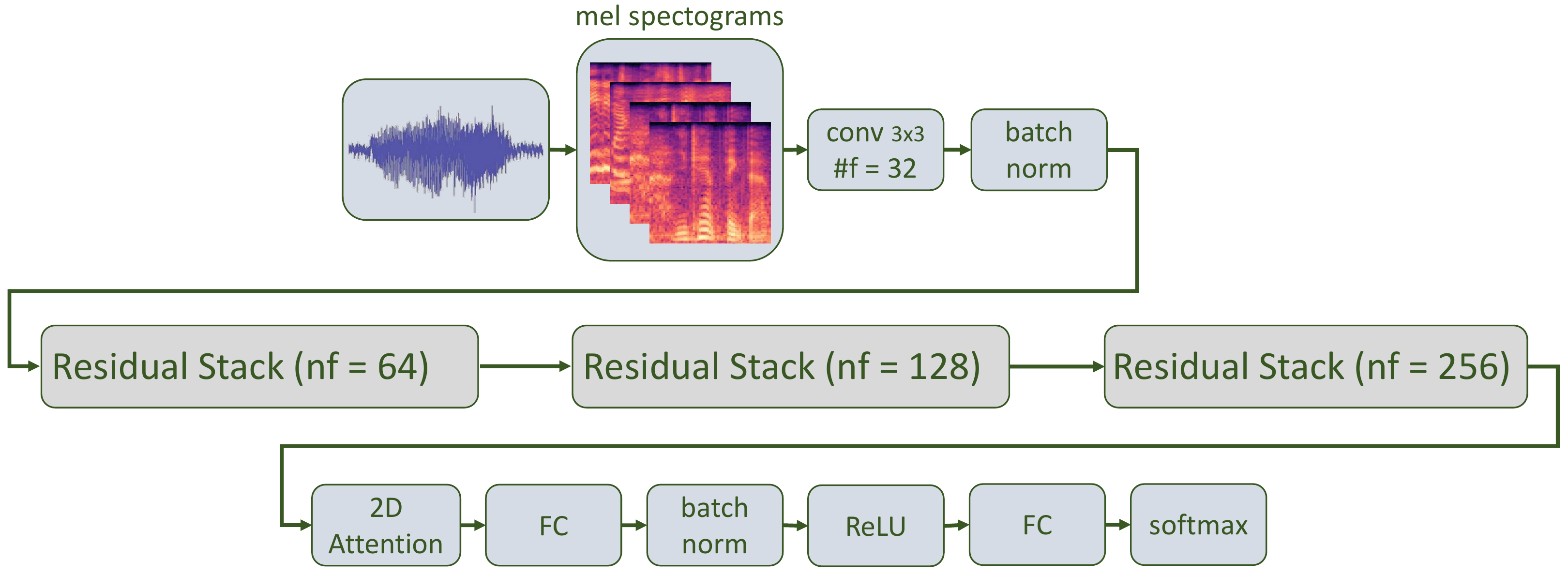}
\end{center}
\caption{Architecture of the base ResNet model used in the experiments for multi-corpus \ac{SER}. Three convolutional stacks extract features from the generated mel-spectrogram input. A 2D attention module is then applied to reduce the variable length output of the convolutional base to a single feature vector for further processing by a \ac{MLP} classifier head. $nf$ specifies the number of filters ($\#f$) of all convolutions inside a specific residual stack.
}
\label{fig:resnet}
\end{figure*}

All ResNet models trained on \emoset contain three submodules (stacks) of residual blocks chained in sequence. Before those submodules, however, an additional convolutional block is applied. This block consists of a convolutional layer with 32 filters of size $3\times3$, which are convolved with the input using a stride of 1. Additionally, \ac{BN} is applied to the outputs of the convolution. Afterwards, the output is fed through the submodules. The number of filters in the convolutional layers within these submodules doubles for each consecutive one. The first one uses 64, the second 128, and the third one 256 filters of shape $3\times3$. Each of the modules contains two residual blocks, where the blocks are structured as follows. In the very first block, a convolutional layer with a stride of 2 is applied to the input followed by \ac{BN} and a \ac{ReLU} activation. A second convolution, this time with a stride of 1, is placed directly afterwards, its output again batch normalised. As the number of filters in the block's first convolution increases compared to the input received from the very first convolution in the network (from 32 to 64), a shortcut connection is needed to add the residual (input to the block) back to the output received after the two convolutions. This is done by first applying average pooling with patches of $2\times2$ and stride 2 to the block input and concatenating this pooled residual with zeroes along the filter dimension. The resulting residual is then added to the output of the second convolution in the block and a \ac{ReLU} activation is applied. The second block differs only in that it uses a stride of 1 for both of its convolutional layers and further has no need for the shortcut connection as the number of filters does not increase. For the other two submodules, it works the same but the number of filters are increased to $128$ and $256$, respectively. As for the first module, only the first block in each of the other two modules has convolutions with stride 2 and needs a shortcut connection. A final \ac{BN} and \ac{ReLU} activation are placed after the last block of the third module.

Instead of the standard global pooling and fully connected layers applied on top of the convolutional feature extractor base, a 2D self attention layer as proposed in~\cite{zhao2018deep,zhao2019exploring} is utilised. This layer projects the variable size spatio-temporal output of the \ac{CNN} to a fixed length weighted sum representation. By doing so, the weights are learnt based on the emotional content that can be found in a specific time-frequency region of the spectrogram.
The computation of the weighted sum works as follows (adapted from~\cite{zhao2019exploring}): The output of the convolutional ResNet feature extractor base is three dimensional with size $N_f \times N_t \times N_c$,  where $N_f$ is the number of frequency bins at the output -- for an input with $64$ mel-filters $8$ bins remain after the pooling operations of the chosen architecture have been applied. $N_t$ is the number of time-steps in the input spectrogram again downsampled by $8$, while $N_c$ is the number of channels ($256$) of the last convolutional layer in the ResNet. The first two dimensions of this representation are now flattened such that a sequence of $N_c$-dimensional vectors $x_i$ with  length $N = N_f \cdot N_t$ remains: 
\begin{equation}
\centering
    A= \left\{x_1,\dots,x_N \right\}, x_i \in \realnumbers^{C}.
\end{equation}
Each of these vectors is then transformed into an intermediate learnt representation by a fully connected layer with $256$ units and hyperbolic tangent activation. Afterwards, an importance vector is calculated by the inner product of the layer output and a learnable vector $u$, again of size $256$:
\begin{equation}
    \centering
    e_i = u^{T} tanh(Wx_i+b).
\end{equation}
The attention weight for each individual vector is then computed as a smoothed softmax over all $e_i$s:
\begin{equation}
    \centering
    \alpha_i = \frac{\exp{(\lambda e_i)}}{\sum_{k=1}^{N} \exp{(\lambda e_k)}}.
\end{equation}
The factor $\lambda$ defines a smoothing of the attention weights. If $\lambda = 0$, then each of vector gets an equal weight of $1/N$, while with $\lambda = 1$, no smoothing of the computed attention weights is performed. The first case can be considered as a global average pooling the ResNet output. A value of $\lambda = 0.3$ has been found to work well for \ac{SER} on the \ac{IEMOCAP} database by Zhao \etal in~\cite{zhao2018deep, zhao2019exploring}, and is adopted here as well. Finally, the output of the module computes the attention weighted sum of the flattened input vectors $x_{1,\dotso,N}$ as:
\begin{equation}
    \centering
    c = \sum_{i=1}^{N} \alpha_i x_i.
\end{equation}
Similar to the weighted sum of fixed size vectors, $c$ has a fixed size of $256$. Compared to standard global average pooling, the attention layer introduces trainable parameters that aim to learn the importance of a specific time-frequency region's features to the training task at hand. Further, this module allows for training with variable length audio sequences. A stack of fully connected layer, \ac{BN} and dropout is finally put on top of the attention module before the softmax classification layer. The attention layer can contain corpus specific parameters or be shared across datasets while the fully connected stack always serves as a domain specific classifier, \ie, is not shared between databases, such as the adapter modules. \Cref{fig:resnet} gives a visual overview of the whole model.

\begin{figure}[tp]
\begin{center}
\includegraphics[width=0.7\columnwidth]{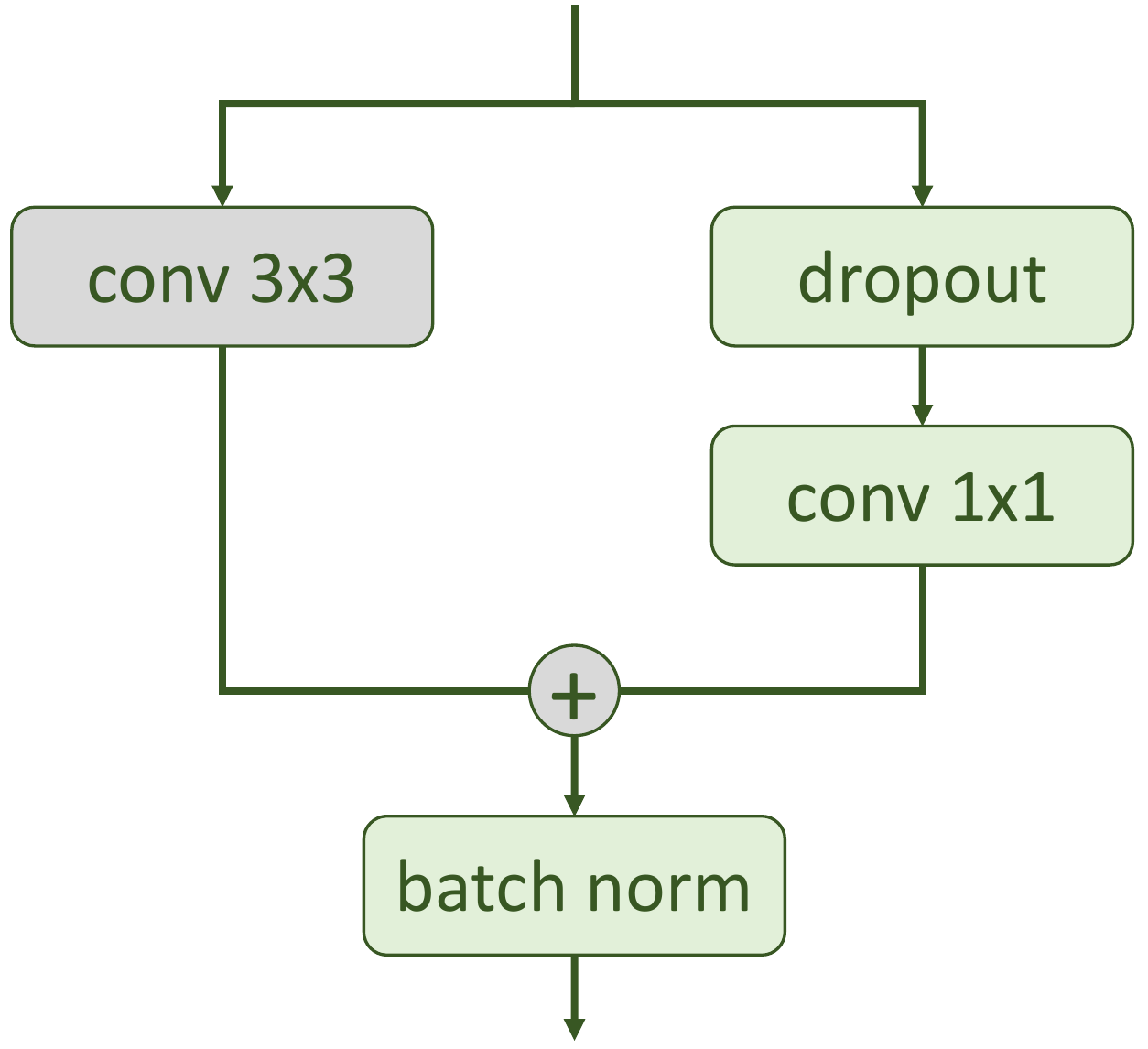}
\end{center}
\caption{Depiction of a residual adapter module. The adapter is a task specific small convolution ($1\times1$) that is applied in parallel to all convolutions of the shared base model. The outputs of both convolutions are then combined by their elementwise summation. Additionally, the subsequent \ac{BN} which is not shared between corpora is shown in the figure.}
\label{fig:res-adapt}
\end{figure}

For transfer learning, residual adapters for each target database are used throughout the whole depth of the network and are applied in parallel to the shared convolutions. The adapter modules are convolutional layers which all contain small $1\times1$ kernels. They share the rest of their settings (strides and number of filters) with the corresponding convolution. The output of each adapter is then added to that of the parallel convolution and fed through the \ac{BN} layer (cf.~\Cref{fig:res-adapt}). Further, all \ac{BN} layers contain parameters which are trained for each database. Depending on the number of neurons in the classification softmax layer, the described architecture has around 3 million parameters in total of which  300\,000 are domain specific.

\subsubsection{Training Procedure}
\label{sssec:resnet-train}
The experiments can be divided into two parts. For the first set of experiments, the transfer capabilities of a ResNet with adapter modules which are trained on a single main task is evaluated for all tasks in \emosetns. All parameters of the base model are trained on either \ac{DEMoS}, \ac{IEMOCAP}, \ac{AIBO}, or \ac{GEMEP}. Subsequently, the pre-trained model is transferred by freezing all parameters apart from the task specific classifier head and adapters. Afterwards, the remaining parameters are trained on the new task.
The second set of experiments takes multiple corpora into account for training the base model. Here, a model is constructed which includes adapter modules and classifier heads for each of the training tasks while sharing the rest of the parameters. After the training process is finished,
the shared parameters are frozen and transferred to a new task from \emoset, while adapters and classifier are reinitialised.

For both kinds of experiments, the following hyperparameters were chosen for the training procedure. The models are optimised via \ac{SGD} with momentum of $0.9$ in batches of $64$ examples. Further, as \ac{BN} is used throughout the model, the initial learning rate is set to a high value of $0.1$ and exponentially decayed by a small factor of $10^{-6}$.
Finally, the learning rate is reduced in three steps from $0.1$ to $0.01$ and ultimately to $0.001$.
For the single-task transfer experiments and training the baseline models, training is halted and continued with a smaller learning rate if no increase in validation \ac{UAR} has occurred for 50 epochs of training. While these parameters are applied for both adapter tuning and training from scratch on each of the corpora, full finetuning and tuning only the classifier head start training with a reduced learning rate of $0.01$.

When training on multiple datasets at once, a round-robin approach is applied, sampling one batch of each dataset. One round is further considered as one step when defining the learning rate schedule. After a fixed number of $2\,500$ round-robin steps, the learning rate is set to the according value. After the second step, training continues for an additional $2\,500$ steps and then stops. After the shared model training, the adapters and classifier heads are further finetuned for each task in the same way as described above for the single-task transfer experiment. This additional finetuning step should help individual models which did not reach their global optimum at the very end of the shared model training procedure.

\section{Cross-Corpus Strategies}
\label{sec:cross-corpus-strategies}
For training the proposed deep learning models on the \emoset corpus, two general directions are followed.
First, a (mostly-)shared model can be trained jointly on all corpora in a multi-domain/task fashion. Second, the individual emotion classification corpora can be aggregated into a single large corpus by mapping the different available emotion categories into a shared label space.

\subsection{Shared Model Multi-Domain Training}
\label{ssec:multi-domain}
For training a shared model with multiple classifier heads on the \emoset databases simultaneously, batches of samples from the individual datasets are sampled sequentially in a round-robin fashion, only updating parameters specific to this dataset or belonging to the shared model.
In addition to the classifier heads, corpus specific residual adapter modules are introduced in parallel to the shared network's convolutions. In practice, this means that a training batch belonging to a specific \emoset domain is used to update all of the shared model's parameters in addition to the respective domain specific adapters and classifier head. 

\footnotesize
\begin{table*}
\footnotesize
\centering
\caption{Mapping of classes contained inside the \emoset databases onto six classes of arousal and valence combinations. Categories are mapped as eliciting either low or high arousal. For valence, negative, neutral, and positive emotions are categorised.}
\label{tab:AV}
\begin{tabular}[t]{llllll}

\toprule
\multicolumn{3}{c}{\textbf{low A}} & \multicolumn{3}{c}{\textbf{high A}} \\
                     \textbf{negative V} &  \textbf{neutral V} & \textbf{positive V} & \textbf{negative V} & \textbf{neutral V} & \textbf{positive V} \\
\midrule

  \makecell[tl]{contempt\\disappointment\\disgust\\frustration\\guilt\\hurt\\impatience\\irritation\\jealousy\\sadness\\shame\\unfriendliness\\worry} & \makecell[tl]{boredom\\confusion\\neutral\\pondering\\rest\\sneakiness} & \makecell[tl]{admiration\\kindness\\pride\\relief\\tenderness} & \makecell[tl]{aggressiveness\\anger\\anxiety\\despair\\fear\\helplessness\\high-stress\\scream} & \makecell[tl]{emphatic\\interest\\intoxication\\medium-stress\\nervousness\\surprise} & \makecell[tl]{amusement\\cheerfulness\\elation\\excitement\\happiness\\joking\\joy\\pleasure\\positive} \\ \hline
\end{tabular}
\end{table*}
\normalsize

\subsection{Aggregated Corpus Training via Arousal Valence Mapping}
\label{ssec:aggregated}
The second way of training in the multi-corpus \ac{SER} setting considers ways of aggregating the individual corpora into a larger shared database. As all databases deal with emotion classification, a suitable approach is to map the annotated emotion categories into a shared label space for training. A fitting mapping is to use a dimensional approach and transform the multi-class problems into binary and three-way arousal/valence 
classification. As described in~\Cref{sec:ser}, a method for this mapping is to use a model, such as the circumplex of affect~\cite{russell1980circumplex}. In \Cref{tab:AV}, such a mapping has been performed for the emotions included over all of \emosetns's databases.
In order to prevent the model from learning the class distributions of the \emoset corpora, random subsampling
is performed on a per dataset basis. For each corpus, the number of samples for each of the mapped classes, \ie, the two arousal and three valence levels, is equal to the respective sample count of the minority class. This is applied on training, validation, and testing partitions of each dataset separately. Training with the aggregated corpus approach can then proceed as in standard single dataset network training, considering samples from all datasets as input and learning to predict the 
arousal or valence categories of the joint mapped label space. Finally, a combination of the multi-domain and the aggregated corpus approach can be taken for training a single network on both arousal and valence mapped corpora, \ie, the domains are represented by the respective arousal and valence aggregated corpora. 

\section{Evaluation}
\label{chap:evaluation}
The results for all baselines and each dataset in \emoset are briefly analysed in~\Cref{sec:baseline-results}. Afterwards, results of the transfer learning experiments with the residual adapters (in~\Cref{sec:eval-res-adapt}), are discussed.

\subsection{Baseline Results}
\label{sec:baseline-results}

\begin{table*}
\centering
\caption{Baseline results in \ac{UAR} for the different approaches: A ResNet trained on each \emoset database individually, an eGeMAPS + SVM and a \ds + SVM baseline. For the ResNets trained from scratch, two different sets of results are presented: First, the evaluation results for the best training epoch (measured in development \ac{UAR}) can be found in the first columns. Secondly, the results achieved at the very end of training are reported in the last columns. Further, for each dataset the chance level is given (in \ac{UAR}.}

\footnotesize
\label{tab:baseline}
\begin{tabular}{lrrrrrrrrr}

\toprule
{[\,\%]} &  & \multicolumn{4}{c}{ResNet} & \multicolumn{2}{c}{eGeMAPS} & \multicolumn{2}{c}{DS} \\
{} & {} & \multicolumn{2}{c}{best} & \multicolumn{2}{c}{final} & \multicolumn{2}{c}{}\\
Dataset & chance & devel & test & devel & test & devel & test & devel & test \\
\midrule
\ac{ABC} & 25.0     &    44.9 &         41.9 &                42.7 &               39.4 & \textbf{47.0} & \textbf{54.4} & 45.2 & 33.5 \\
\ac{AD}   &  50.0     &          86.1 &         71.4 &                82.4 &               72.9 & \textbf{88.5} & \textbf{78.7} & 82.5 & 71.9 \\
\ac{BES}  &   16.7    &   \textbf{65.0} &         55.0 &                63.3 &               53.3 & 53.3 & 58.3 & 55.0 & \textbf{63.3} \\
\acs{CASIA}         &      16.7    &          \textbf{36.3} &         18.7 &                29.3 &               23.3 & 33.7 & 24.7 & 26.3 & \textbf{31.3}  \\
\ac{CVE}    &    14.3   &   35.6 &         30.3 &                30.9 &               34.2 & \textbf{58.9} & \textbf{60.4} & 48.5 & 52.2 \\
\ac{DEMoS}  &   14.3  &   \textbf{89.0}  &   \textbf{73.8}  &   88.9  &   73.6  &   38.4  &   43.2 & 53.0 & 46.9 \\
\ac{DES}    &   20.0  &   \textbf{34.7}  &   43.3  &   21.9  &   \textbf{52.6}  & 31.0  &   46.5 & 23.0 & 34.4 \\
EA-ACT      &      14.3      &          \textbf{32.9} &         35.7 &                12.9 &               50.0 & 22.9 & \textbf{57.1} & 24.3 & \textbf{57.1} \\
EA-BMW       &       33.3    &          73.4 &         46.7 &                71.4 &               56.3 & \textbf{79.8} & 56.3 & 61.1 & \textbf{59.3}\\
EA-WSJ        &     50.0     &         \textbf{100.0} &         98.1 &               \textbf{100.0} &               98.1 & \textbf{100.0} & \textbf{100.0} & 98.1 & 98.1\\
\ac{EMO-DB}    &        14.3      &          68.8 &         59.2 &                63.5 &               \textbf{61.6} & \textbf{71.9} & 48.4 & 52.5 & 61.3\\
\ac{eNTERFACE}       &   16.7     &          \textbf{71.5} &         81.0 &                71.1 &               \textbf{82.4} & 44.8 & 49.1 & 50.0 & 46.2 \\
\ac{EU-EV}      &    5.6   &          \textbf{12.7} &         10.4 &                 6.3 &               \textbf{11.1} & 7.8 & 10.9 & 8.1 & 11.0 \\
EmoFilm           &   20.0   &          46.3 &         46.3 &                44.4 &               47.0 & \textbf{46.9} & 54.6 & 46.8 & \textbf{54.7} \\
\ac{EmotiW}       &  14.3    &          28.1 &         24.1 &                24.0 &               24.3 & 30.5 & \textbf{30.6} & \textbf{33.3} & 29.2  \\
\ac{AIBO}        &   20.0    &          \textbf{53.3} &         37.2 &                51.7 &               36.9 & 46.2 & \textbf{38.3} & 43.8 & 36.8 \\
\ac{GEMEP}        &      5.9     &          \textbf{46.0} &         29.1 &                44.6 &               29.8 & 41.1 & \textbf{35.4} & 35.2 & 22.4 \\
\ac{GVEESS}        &      7.7     &          26.9 &         24.7 &                17.8 &               20.4 & \textbf{38.5} & 21.2 & 32.7 & \textbf{27.9}\\
\ac{IEMOCAP}    &    25.0     &          52.7 &         53.6 &                49.6 &               57.6 & \textbf{56.2} & \textbf{58.4} & 54.4 & 57.7 \\
\ac{MELD}        &    16.7       &          \textbf{24.8} &         21.7 &                22.4 &               20.2 & 23.7 & \textbf{23.1} & 23.8 & 22.2\\
\ac{MES} &   16.7    &   \textbf{66.7} &         \textbf{75.0} &                56.7 &               70.0 & 61.7 & 61.7  & 38.3 & 43.3 \\
PPMMK        &    25.0   &          \textbf{70.1} &         40.6 &                69.5 &               39.3 & 48.5 & 40.4 & 58.6 & \textbf{40.8} \\
\ac{SIMIS}        &     20.0      &          \textbf{40.7} &         30.5 &                38.7 &               24.7 & 38.5 & \textbf{31.7} & 33.3 & 29.5 \\
\ac{SmartKom}          &   14.3   &          20.2 &         20.8 &                11.3 &               23.3 & 25.2 & \textbf{28.6} & \textbf{25.7} & 26.2 \\
\ac{SUSAS}         &     25.0     &          \textbf{64.2} &         56.5 &                54.2 &               \textbf{59.4} & 56.6 & 56.5 & 54.6 & 59.3\\
TurkishEmo      &  25.0  &          62.5 &         55.7 &                56.8 &               \textbf{58.0} & \textbf{64.8} & 56.8 & 54.5 & 53.4 \\
\bottomrule
\end{tabular}
\end{table*}

Three baseline systems are tested for their efficacy on each corpus of the \emoset separately: i) the eGeMAPS \ac{SVM} combination (cf.~\Cref{ssec:eval-egemaps}), ii) the \ds feature extraction with linear \ac{SVM} (cf.~\Cref{ssec:eval-ds}), and iii) the ResNet architecture trained from scratch on the mel-spectrograms of each dataset (cf.~\Cref{ssec:eval-resnet}).

The results of these methods can be found in~\Cref{tab:baseline}. From a high level view of these results, it can be seen that -- among those baselines -- there is no overall best approach to solving the \ac{SER} classifications for every dataset. Rather, depending on the corpus, the best achieved result can be found in any of the models. Furthermore, results for some corpora fluctuate quite heavily, which becomes especially apparent by looking at the two presented types of results of the ResNet trained from scratch on each corpus. These two points are discussed in the following sections where appropriate.

\subsubsection{eGeMAPS}
\label{ssec:eval-egemaps}
As an expert-designed, handcrafted acoustic feature set specifically intended for paralinguistic speech analysis tasks, \ac{eGeMAPS} is a competitive baseline for many of the included corpora in \emosetns. While it achieves the best results on the test partition of $12$ databases, the margins by which it does differ. Large increases over the other two baselines can be found on \ac{ABC}, \ac{CVE}, and \ac{GEMEP} with a delta of around $10$ percentage points compared to the second best approach. Especially \ac{GEMEP} with its $17$ annotated classes seems to benefit from using a small higher-level feature set in combination with an \ac{SVM} classifier. Slight increases from the other two baselines occur for \ac{EmotiW}, \ac{AIBO}, \ac{IEMOCAP}, \ac{MELD}, \ac{SIMIS}, and \ac{SmartKom}, while performance for the EA datasets (EA-ACT, EA-BMW, and EA-WSJ), \ac{SUSAS}, and Turkish Emo is on par with the other baseline method. Surprisingly, \ds and the ResNet baseline both achieve considerably better results particularly on the test partition of \ac{EMO-DB}.
This might be caused by the optimisation procedure of the baselines utilising \acp{SVM} as classifiers. The complexity parameter is optimised from a fixed logarithmic set of values based on the performance on the development partition, which might not always be the best value to choose for performing the fit on the combined training and development partitions for evaluation on test.

\subsubsection{Deep Spectrum}
\label{ssec:eval-ds}
With a few exceptions, combining \ds features with a linear \ac{SVM} classifier leads to results comparable to the \ac{eGeMAPS} baseline method. Negative examples can be found with \ac{ABC}, \ac{DES}, and \ac{GEMEP}, where the \ds system falls behind both other baselines. On \ac{GVEESS}, on the other hand, it achieves the best test set result of $27.9\,\%$ compared to the runner-up with $24.7\,\%$. For the EA datasets, \ac{EU-EV}, and EmoFilm it matches performance with \ac{eGeMAPS}. In the case of \ac{DEMoS}, the system lags behind the ResNet trained from scratch considerably but improves on the \ac{eGeMAPS} baseline. This might be explained by the larger size of the \ds features which can provide more discriminative features for the \ac{SVM} classifier given enough training samples. When looking at \ac{BES} and \ac{MES} -- two datasets that only differ in the recorded subjects and their spoken language -- it can be seen that the system has problems consistently handling small datasets with a few number of speakers: In the case of \ac{BES}, \ds achieves the best result on the test set of all the investigated baselines, while it is considerably less performant on \ac{MES}. The nature of these datasets also has an impact on the other approaches, but is overall more pronounced here. 

\subsubsection{ResNet from scratch}
\label{ssec:eval-resnet}
Results for the last baseline against which transfer learning experiments with residual adapters are evaluated were obtained by training a ResNet with architecture and training settings as described in~\Cref{sssec:resnet-train} on each individual task. 
A first set of development and test results restore the model which achieved the best \ac{UAR} on the development partition for evaluation on the held-out test set. For the second type of results
, the development and test set \acp{UAR} achieved at the very end of the training are reported. The reason for reporting both of those sets of values is motivated by the nature of the training process and the databases. Training of the networks, as described in~\Cref{sssec:resnet-train} always starts with a high learning rate which is reduced in steps after validation performance has not increased for a certain patience period. As many datasets contain only a few hundred samples, this can lead the model to explore into an area of the learning landscape which -- by chance -- results in very good results on the development partition that do not correspond to similar results on the test partition very early in the training process. If by the end of training, better performance on the development partition has not been achieved, this very early model is used for the final evaluation on the test partition.

All results can be found alongside the other two baselines in~\Cref{tab:baseline}.
In the other two baseline approaches, the usage of a linear \ac{SVM} has an advantage over the neural network approach, in that the combined training and validation data can be used to perform a final fit of the classifier after having found the optimal training parameters. In the ResNet experiments, however, the validation data is used to measure the model's performance and generalisation capabilities during training. Furthermore, with the help of the validation data learning rate adjustments are defined and the training is stopped early before performing the final evaluation on the held-out test set.
Thus, the samples in the validation partition have no immediate impact on the weights of the trained model which might limit the generalisation capabilities especially for datasets containing only a few number of samples or speakers. 

As the validation and testing splits contain different sets of speakers, cases exist where validation and testing performance diverge heavily. For example, for \ac{DES}, the model achieves a test set \ac{UAR} of $43.3\,\%$ against the overall best development performance of $34.7\,\%$, but at the end of training, this discrepancy increases to $52.6\,\%$ on test, and only a mere $21.9\,\%$ on development which is near chance-level. A very similar picture is observed for EA-ACT, where development and test performance diverge to $12.9\,\%$ against $50.0\,\%$. The performance on these datasets is further hindered by the fact that they only contain a small number of training and validation samples. Also, a few of the datasets can be identified as being challenging for the ResNet model in general, such as \ac{EU-EV}, which is a corpus with a very large number of annotated emotion classes in three different languages. In addition, the training and evaluation setup chosen for \emoset once more increases the difficulty by partitioning based on language. In the end, \acp{UAR} of around $10\,\%$ are achieved by the model if only the corpus itself is used for training which is in line with the other baselines. For \ac{EmotiW}, the challenge lies in its multi-modal nature. The corpus additionally contains visual content,  which is immensely helpful in the identification and discrimination of emotions through the analysis of facial characteristics. As could already be seen in the baseline using eGeMAPS features and a linear \ac{SVM} classifier, information extracted from only the audio content does not lead to favourable results for this corpus. On both \ac{SIMIS} and \ac{SmartKom},  the ResNet achieves only very weak results which are slightly above chance-level. Here, \ds and eGeMAPS perform better. For eNTERFACE on the other hand, this approach substantially outperforms the \ds system reaching a test set \ac{UAR} of over $80\,\%$. Furthermore, for this dataset, the \acp{UAR} at the best and final model checkpoint are consistent, \ie, the model converged to an optimum. Especially weak performance can be found on~\ac{CVE}, with the ResNet trained from scratch falling behind the other two baselines by more than $20\,\%$ \ac{UAR} when compared to \ac{eGeMAPS}, and $15\,\%$ against \ds on both development and test partitions. Having only 4 speakers in total, this is another dataset where the danger of overfitting to the training data is especially high for a deep learning model that is trained from the raw audio content. Here, the \ac{eGeMAPS} and \ds have the advantage of utilising abstracted feature representations in the form of an expert-designed, hand-crafted audio descriptor set and high-level image representations learnt from the task of object recognition. The strongest ResNet result is achieved for \ac{DEMoS} with a test set \ac{UAR} of $73.8\,\%$ which is around $30\,\%$ above the other baseline methods. This can be explained by the partitioning of the dataset for \emoset which has separated non-prototypical from prototypical emotion portrayals in a speaker dependent way. As the ResNet approach learns most directly from the raw input data, fine-grained emotionally discriminative features for each speaker can be learnt from the low-level spectrogram representation. Adding layers of abstraction to the feature representation in the \ac{eGeMAPS} and \ds baselines hides away this information.

\begin{figure*}[tph]
\begin{subfigure}[t]{.49\textwidth}
\raisebox{-0.005\height}{\includegraphics[height=0.806\textheight]{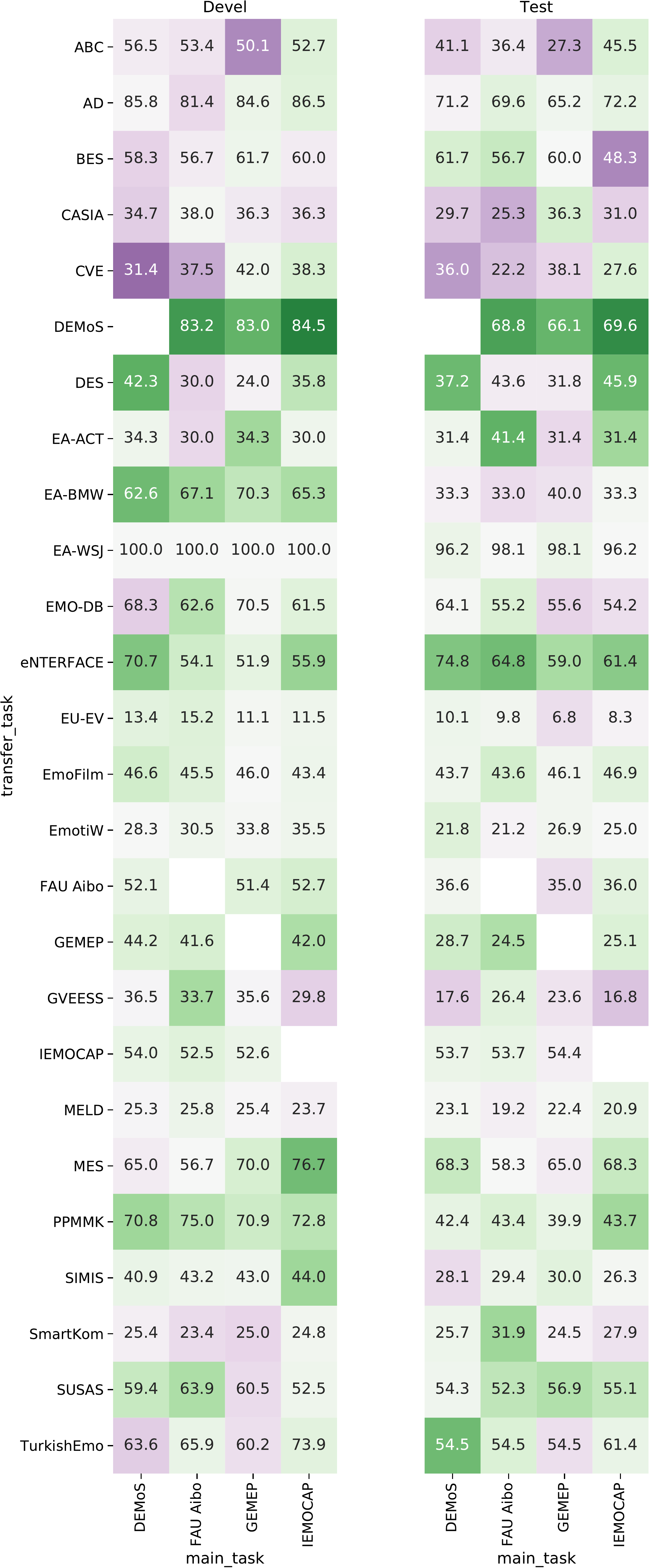}}
\caption{\acp{UAR} achieved on the development and test partition of each \emoset corpus when performing adapter transfer learning from a ResNet pre-trained on one of four main tasks. The colours represent increases and decreases from classifier head tuning.}
\label{fig:single-task/cnn/adapters-last}
\end{subfigure}
\hfill
\begin{subfigure}[t]{.49\textwidth}
\includegraphics[height=0.8\textheight]{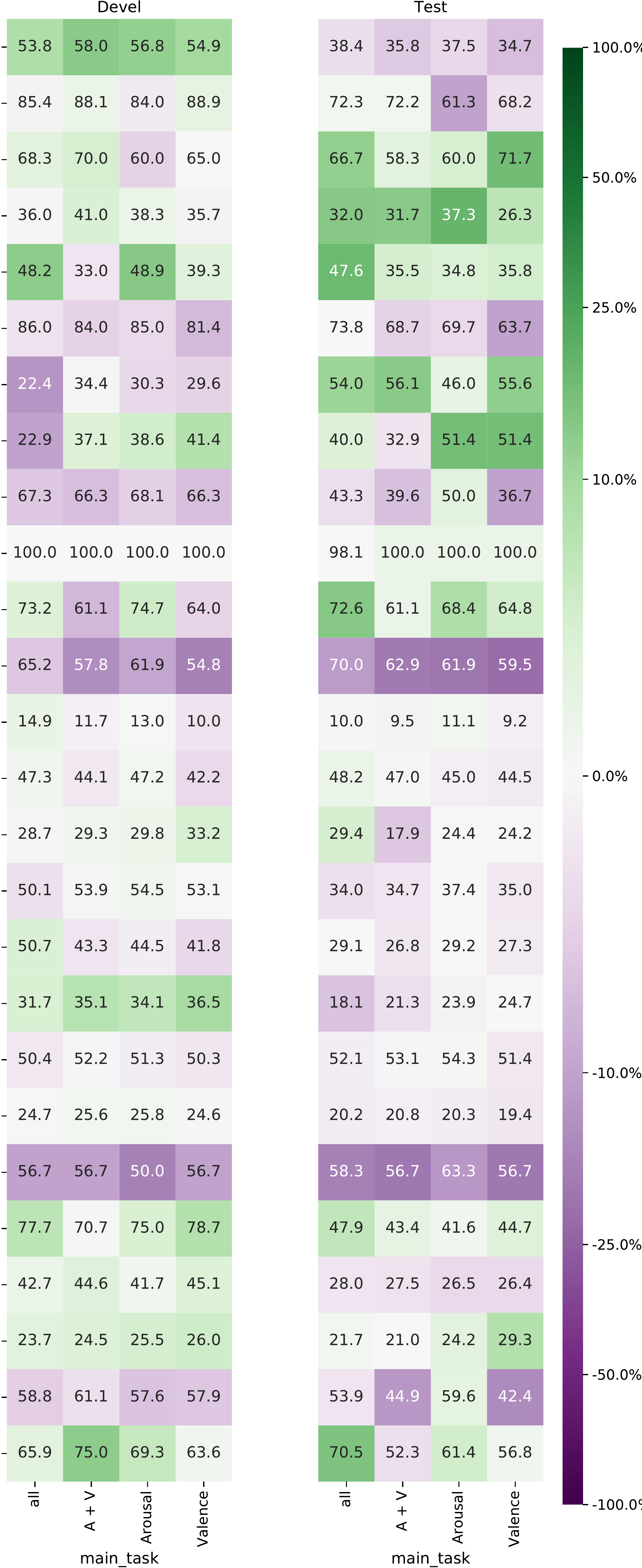}
\caption{\acp{UAR} achieved at the best epoch on the development and test partition of each \emoset corpus when performing adapter transfer learning from a ResNet with a shared 2D attention module pre-trained in different multi-task settings on the \emoset corpora. The colours of the heatmap represent the increase or decrease from the \acp{UAR} achieved by a model trained solely on the target corpus percentage points.}
\label{fig:multi-task/cnn/sfl-True}
\end{subfigure}
\end{figure*}

\begin{table*}
\centering

\caption{Comparison of achieved test set results in the transfer learning experiments with residual adapters utilising all of \emosetns. ``All'' denotes training a single multi-corpus model wich classifier-heads for every database in a round-robin fashion. McNemar's test is used to test for statistical difference ($p<0.05$) of the proportion of errors made by the transfer models compared to the baseline. ``$^{+}$'' denotes a statistically significant improvement, ``$^{-}$'' a decrease. Note that a performance increase according to the McNemar test does not correlate with a higher \ac{UAR} for very imbalanced datasets.}
\footnotesize
\label{tab:adapter_transfer}
\resizebox{.6\textwidth}{!}{
\begin{tabular}{lllllll}
\toprule
{[\,\%]} &  \\

Dataset & chance & baseline & all & Arousal & Valence & A + V  \\
\midrule
\ac{ABC} & 25.0     &    \textbf{41.9} & 38.4  & 37.5 & 34.7 & 35.8 \\
\ac{AD}   &  50.0     &  71.4 & \textbf{72.2} & $61.3^{-}$ & 68.2 & \textbf{72.2} \\
\ac{BES}  &   16.7    &   55.0 & $66.7^{+}$ & 60.0 & $\textbf{71.7}^{+}$ & 58.3 \\
\acs{CASIA}         &      16.7    &   18.7 & $32.0^{+}$ & $\textbf{37.3}^{+}$ & $26.3^{+}$ & $31.7^{+}$ \\
\ac{CVE}    &    14.3   &   30.3 & $\textbf{47.6}^{+}$ & $34.8^{+}$ & $35.8^{+}$ & $35.5^{+}$ \\
\ac{DEMoS}  &   14.3 &   73.8 & \textbf{73.9} & $69.7^{-}$ & $63.7^{-}$ & $68.7^{-}$  \\
\ac{DES}    &   20.0  &   43.3 & 54.0 & 46.0 & $\textbf{55.6}^{+}$ & $56.1^{+}$ \\
EA-ACT      &      14.3 &  35.7 & 40.0 & 51.4 & $\textbf{51.4}^{+}$ & 32.9 \\
EA-BMW       &       33.3    &   46.7 & 43.3 & \textbf{50.0} & 36.7 & 39.6\\
EA-WSJ        &     50.0 &         98.1 &  98.1 & \textbf{100} & \textbf{100} & \textbf{100} \\
\ac{EMO-DB}    &        14.3 & 59.2 & $\textbf{72.6}^{+}$ & $68.4^{+}$ & 64.8 & 61.1 \\
\ac{eNTERFACE}       &   16.7     & \textbf{81.0} & $70.0^{-}$ & $62.0^{-}$ & $59.5^{-}$ & $61.4^{-}$ \\
\ac{EU-EV}      &    5.6  & 10.4 & $\textbf{11.1}^{+}$ & 9.1 & 9.2 & $9.5^{-}$\\
EmoFilm           &   20.0 &  46.3 & \textbf{48.2} & 45.0 & 44.5 & 47.0\\
\ac{EmotiW}       &  14.3  & 24.1 & \textbf{29.4} & 24.4 & 24.2 & $17.9^{-}$ \\
\ac{AIBO}        &   20.0 & 37.2 & $34.0^{-}$ & $\textbf{37.4}^{-}$ & $35.1^{-}$ & 34.7 \\
\ac{GEMEP}        & 5.9 & 29.1 & 29.1 & \textbf{29.2} & 27.3 & 26.8 \\
\ac{GVEESS}        &      7.7     &  24.7 & 18.1 & 23.9 & \textbf{24.7} & 21.3 \\
\ac{IEMOCAP}    &    25.0  & 53.6 & 52.1 & \textbf{54.4} & $51.4^{-}$ & $53.1^{-}$\\
\ac{MELD}        &    16.7  &  \textbf{21.7} & $20.2^{+}$ & $20.3^{+}$ & $19.4^{+}$ & $20.7^{+}$ \\
\ac{MES} &   16.7    &  \textbf{75.0} & $58.3^{-}$ & $63.3^{-}$ & $56.7^{-}$ & $56.7^{-}$ \\
PPMMK        &  25.0  &  40.6 & $\textbf{47.8}^{+}$ & 41.6 & 44.7 & 43.4 \\
\ac{SIMIS}        &     20.0 & \textbf{30.5} & $28.0^{-}$ & $26.5$ & 26.4 & $27.5^{-}$ \\
\ac{SmartKom}          &   14.3  & 20.8 & $21.7^{+}$ & $24.2^{+}$ & $\textbf{29.3}^{+}$ & 21.0 \\
\ac{SUSAS}         &     25.0  & 56.5 & $53.9^{-}$ & \textbf{59.6} & $42.4^{-}$ & $44.9^{-}$ \\
TurkishEmo      &  25.0 & 55.7 & $\textbf{70.5}^{+}$ & 61.4 & 56.8 & 52.3 \\
\bottomrule
\end{tabular}
}
\end{table*}

\subsection{Parallel Residual Adapters}
\label{sec:eval-res-adapt}
In the case of the parallel residual adapter models trained on \emosetns, experiments and their evaluation are proceeded in a slightly different way. In a first set of experiments, denoted as ``single-task transfer'', four tasks out of \emoset were chosen as base tasks for pre-training the deep learning architectures while for ``multi-task transfer'', all \emoset corpora are used to pre-train a shared model. In both cases, afterwards, a transfer experiment is run for each \emoset task by training only the adapter modules on the individual tasks data. Finally, the performance of the transfer learning approach is evaluated by comparing the achieved test set \ac{UAR} to that of a model with same architecture trained from scratch on the specific task, and to more traditional transfer approaches, \ie, full finetuning and classifier head tuning.
\subsubsection{Single-Task Transfer}
\label{ssec:single-task-transfer}
The \emoset corpora \ac{DEMoS}, \ac{AIBO}, \ac{GEMEP}, and \ac{IEMOCAP} are chosen as base pre-training tasks based on a number of qualitative and quantitative characteristics: With \ac{DEMoS}, \ac{IEMOCAP}, and \ac{GEMEP}, three acted \ac{SER} corpora are included which contain either a large amount of training samples (\ac{DEMoS}, and \ac{IEMOCAP}) or annotated classes (\ac{GEMEP}).

For brevity, we leave out the detailed results of these transfer experiments and only summarise our findings. When comparing against a ResNet that is trained on the target corpus from scratch, adapter tuning from a model trained on a single source corpus achieved mixed results.

While for some, especially smaller tasks, such as \ac{BES} or \ac{CASIA} increases in \ac{UAR} could be observed, in most cases, performance on both development and test partitions stayed roughly the same. A noteworthy negative example was \ac{eNTERFACE}, for which the performance drops noticeably compared to training a full model on only the target data. A model trained exclusively on this dataset is able to achieve a \ac{UAR} of $81\,\%$ while the best result for the transfer experiments was more than 5 percentage points below that at $75\,\%$ \ac{UAR}. On the other hand, \ac{SmartKom}, a large but difficult corpus of natural emotional speech,  seemed to benefit from pre-training and adapter transfer in all cases.
Nevertheless, these results were encouraging when observed from another perspective.  They showed that the features learnt by the ResNet model from Mel-spectrograms of one \ac{SER} corpus can be used reasonably well for a large range of \ac{SER} tasks by simply introducing a small amount of additional parameters -- the adapter modules. 

In addition to comparisons against training models on each task from scratch, the residual adapters approach should be related to more traditional finetuning strategies. The pre-trained models can be taken as feature extractors and only the last classification layers are re-trained for each task. This method tunes an even smaller amount of parameters for each corpus, further decreasing the risk of overfitting but having decreased learning capabilities. 

A comparison between adapter and classifier head tuning is made in~\Cref{fig:single-task/cnn/adapters-last}. Apart from very few exceptions, it can be seen that adapter tuning beats the feature-extraction transfer approach in terms of \ac{UAR} both on the development and test partitions. Notable outliers can be found with \ac{CASIA} and \ac{CVE} hinting at possible overfitting with the adapters approach. Moreover, for datasets with a large number of classes, \eg, \ac{GVEESS} or \ac{EU-EV} performance on test can vary greatly from run to run, leading to classifier head tuning sometimes outperforming the adapter tuning.

\subsubsection{Multi-Task Transfer}

The second set of experiments using the residual adapters train a shared base network for all \emoset datasets while only the adapter modules and final classification layers are specific to each dataset. In this respect, every task has influence on the weights of the shared feature extraction base while still containing task specific parameters to account for inter-corpus variance. As sharing the 2D attention layer between different corpora was found to have only a minor impact on performance in the initial single-task transfer experiments (cf.\ \Cref{ssec:single-task-transfer}), for multi-domain training, this module is further contained in the shared base model. Here, the performance differences between models trained from scratch and adapted from a pre-trained shared model are further evaluated by performing a McNemar's test~\cite{mcnemar1947note}. Statistically significant ($p<0.05$) differences in the proportion of errors are marked in~\Cref{tab:adapter_transfer}. It should be noted that, due to the highly unbalanced nature of some included databases, a performance increase as measured by the test does not necessarily correspond to a higher achieved test set \ac{UAR}. In the following, whenever differences in the results are described as significant, they are so at $p<0.05$.

The results for multi-task transfer experiments with a ResNet architecture are visualised in~\Cref{fig:multi-task/cnn/sfl-True}.
Four different settings are analysed: First, training on all corpora in a multi-domain setting as described in~\Cref{ssec:multi-domain}, then, the other three settings are given by training on both the aggregated arousal and valence mapped data, either together (\emph{A+V}) or separately (\emph{A} and \emph{V}). Apart from a couple of outliers, model performance for all of \emosetns's corpora either increases or stays the same compared to training a full ResNet model for every dataset when using the adapter transfer approach. Two negative examples are \ac{DES} and EA-ACT, where performance on the development partition only ever slightly increases above chance level. Contrary to this, the performance on the test partition shows an increase for these two datasets that is significant in the case of using the valence pre-trained model. 
As it is already evident in the baseline results, this behaviour is most likely due to the corpora only containing a small number of samples and \ac{DES}'s validation and test partition only containing one speaker each, leading to diverging results. Again, \ac{CASIA} and \ac{CVE} seem to benefit from the transfer learning approach, leading to increases in both development and test set \acp{UAR} from their near chance level performance when training a full ResNet model on their corpus data alone. For the particularly popular baseline \ac{SER} corpus \ac{EMO-DB}, performance is increased on both the development and test by around $10\,\%$, the same is true for the Turkish emotional speech database. Both of these increases are further statistically significant measured by a McNemar's test.

For the choice of training data, training on all of \emoset in a round-robin fashion seems to lead to the best results on both development and test partitions. However, it seems to be closely followed by training the model on the aggregated arousal data alone -- suggesting the features learnt from discriminating arousal across corpora can be effectively tuned for various \ac{SER} tasks with the help of residual adapter modules.
As arousal is generally easier to detect from audio recordings of speech than valence, which is more effectively conveyed and perceived from visual information, such as gestures and mimic, training on the three class valence problem might have been too difficult for the network, thus not leading to emotionally salient feature representations. 
Further, when combining arousal and valence aggregated corpora for training, the individual strengths of pre-training on either corpus do not seem to be complementary.

As evident from~\Cref{tab:adapter_transfer}, the multi-task transfer experiments lead to increased results for 21 of the 26 databases included in \emosetns. For 10 corpora, some of these increases are further statistically significant. Only for 2 databases -- \ac{eNTERFACE} and \ac{MES} -- results are always significantly worse. For \ac{eNTERFACE} this can be explained by the strong performance achieved by a model trained on the corpus from scratch which also beats all of the other considered baselines (cf.~\Cref{tab:baseline}).

\section{Conclusion and Outlook} \label{chap:conclusion}
In this manuscript, we presented a novel deep learning based, multi-corpus \ac{SER} framework -- \emonet -- that makes use of state-of-the-art deep transfer learning approaches. We publicly release the framework including a flexible command line interface on GitHub, such that interested researchers can use it in their own work for a variety of multi-corpus speech and audio analysis tasks. \emonet was investigated and evaluated for the task of multi-corpus \ac{SER} on a  collection of 26 existing \ac{SER} corpora. \emonet adapts the residual adapter approach for multi-domain visual recognition was to the task of \ac{SER} from mel-spectrograms.

Needing only a small portion of additional parameters for each database, it allowed for effective multi-domain training on \emosetns, leading to favourable results when compared to training models from scratch or adapting only the classifier head to each corpus. When all of \emoset is utilised for training, either in a multi-domain fashion or by aggregating the corpora by mapping the included categories to arousal and valence classes, test set performance increases could be achieved for 21 of the 26 corpora. For ten databases, these improvements were statistically significant while on the other hand there are only two datasets (\ac{eNTERFACE} and \ac{MES}) that seem to always be negatively affected by the approach. Compared to fully finetuning a pre-trained model (which requires around ten times the number of trainable parameters), the adapter approach often came out on top. 

The results with utilising the residual adapter model for transfer and multi-domain learning in \ac{SER} motivates further research and exploration. One limitation of the work presented herein is that the base architecture of the ResNet has not been extensively optimised for a speech recognition task. Here, different configurations and variations, \eg, with the number of filters, depth and width of the network, should be evaluated. Moreover, for purposes of constraining computational and time requirements in favour of exploring a wider range of transfer learning settings, the model was kept quite small. Increasing the model size could further improve performance but would require adding a larger amount of training data. This training data could come from large scale audio recognition databases that are not immediately related to \ac{SER}, such as AudioSet~\cite{gemmeke2017audio} or the large scale speaker recognition dataset VoxCeleb~\cite{nagrani2017voxceleb}. Having found an optimised model architecture for training, improvements could further be made by experimenting with the degree of influence each \emoset corpus has on the shared model weights during training. This could for example be investigated by adjusting the probability of sampling a batch from a specific dataset compared to the default round-robin strategy utilised in this manuscript. For the different problem of multi-lingual large-scale \ac{ASR}, residual adapters and probabilistic sampling have been explored in combination with an \ac{RNN} architecture trained on Mel-spectrogram input~\cite{kannan2019large}. As \acp{RNN} are a popular choice for \ac{SER}~\cite{mirsamadi2017automatic, tzirakis2018end}, evaluating the residual adapter approach with these networks in a multi-corpus training setting should be considered. Furthermore, both \acp{CNN} and \acp{RNN} could be modified with adapter modules and then trained simultaneously, combining their high level feature representations. For single-corpus \ac{SER} without adapter modules this has been done in~\cite{zhao2019exploring} with state-of-the-art results. Moreover, so far only \ac{SER} corpora with categorical labels have been considered. Using a multi-domain learning model based on residual adapters, adding databases that are labelled with the dimensional approach and pose regression problems would be possible to further increase the size of the training data. Finally, transferring knowledge between different domains of paralinguistic speech recognition, \eg, the detection of deception from speech, with the help of the adapter approach can be investigated.


\ifCLASSOPTIONcompsoc
  \section*{Acknowledgments}
\else
  \section*{Acknowledgment}
\fi

This research was partially supported by Deutsche Forschungsgemeinschaft (DFG) under grant agreement No.\ 421613952 (ParaStiChaD), and Zentrales Innovationsprogramm Mittelstand (ZIM) under grant agreement No.\ 16KN069455 (KIRun).

\ifCLASSOPTIONcaptionsoff
  \newpage
\fi



\bibliographystyle{IEEEtran}
\bibliography{refs}
%



%

\begin{acronym}
\acro{ABC}[ABC]{Airplane Behaviour Corpus}
\acro{AD}[AD]{Anger Detection}
\acro{AFEW}[AFEW]{Acted Facial Expression in the Wild)}
\acro{AI}[AI]{Artificial Intelligence}
\acro{ANN}[ANN]{Artificial Neural Network}
\acro{ASR}[ASR]{Automatic Speech Recognition}

\acro{BN}[BN]{batch normalisation}
\acro{BiLSTM}[BiLSTM]{Bidirectional Long Short-Term Memory}
\acro{BES}[BES]{Burmese Emotional Speech}
\acro{BoAW}[BoAW]{Bag-of-Audio-Words}
\acro{BoDF}[BoDF]{Bag-of-Deep-Feature}
\acro{BoW}[BoW]{Bag-of-Words}

\acro{CASIA}[CASIA]{Speech Emotion Database of the Institute of Automation of the Chinese Academy of Sciences}
\acro{CVE}[CVE]{Chinese Vocal Emotions}
\acro{CNN}[CNN]{Con\-vo\-lu\-tion\-al Neural Network}

\acro{CRNN}[CRNN]{Con\-vo\-lu\-tion\-al Recurrent Neural Network}

\acro{DEMoS}[DEMoS]{Database of Elicited Mood in Speech}
\acro{DES}[DES]{Danish Emotional Speech}
\acro{DNN}[DNN]{Deep Neural Network}

\acro{eGeMAPS}[eGeMAPS]{extended version of the Geneva Minimalistic Acoustic Parameter Set}
\acro{EMO-DB}[EMO-DB]{Berlin Database of Emotional Speech}
\acro{EmotiW}[EmotiW 2014]{Emotion in the Wild 2014}
\acro{eNTERFACE}[eNTERFACE]{eNTERFACE'05 Audio-Visual Emotion Database}
\acro{EU-EmoSS}[EU-EmoSS]{EU Emotion Stimulus Set}
\acro{EU-EV}[EU-EV]{EU-Emotion Voice Database}

\acro{AIBO}[FAU Aibo]{FAU Aibo Emotion Corpus}
\acro{FCN}[FCN]{Fully Convolutional Network}
\acro{FFT}[FFT]{fast Fourier transform}

\acro{GAN}[GAN]{Generative Adversarial Network}
\acro{GEMEP}[GEMEP]{Geneva Multimodal Emotion Portrayal}
\acro{GRU}[GRU]{Gated Recurrent Unit}
\acro{GVEESS}[GVEESS]{Geneva Vocal Emotion Expression Stimulus Set}

\acro{IEMOCAP}[IEMOCAP]{Interactive Emotional Dyadic Motion Capture}

\acro{LSTM}[LSTM]{Long Short-Term Memory}
\acro{LLD}[LLD]{low-level descriptor}

\acro{MELD}[MELD]{Multimodal EmotionLines Dataset}
\acro{MES}[MES]{Mandarin Emotional Speech}
\acro{MFCC}[MFCC]{Mel-Frequency Cepstral Coefficient}
\acro{MIP}[MIP]{Mood Induction Procedure}
\acro{MLP}[MLP]{Multilayer Perceptron}

\acro{ReLU}[ReLU]{Rectified Linear Unit}
\acro{RMSE}[RMSE]{root mean square error}
\acro{RNN}[RNN]{Recurrent Neural Network}
\acrodefplural{RNN}[RNNs]{Recurrent Neural Networks}

\acro{SER}[SER]{Speech Emotion Recognition}
\acro{SGD}[SGD]{Stochastic Gradient Descent}
\acro{SVM}[SVM]{Support Vector Machine}
\acro{SIMIS}[SIMIS]{Speech in Minimal Invasive Surgery}
\acro{SmartKom}[SmartKom]{SmartKom Multimodal Corpus}
\acro{SUSAS}[SUSAS]{Speech Under Simulated and Actual Stress}

\acro{UAR}[UAR]{Unweighted Average Recall}
\acro{WSJ}[WSJ]{Wall Street Journal}
\end{acronym}
\onecolumn








\end{document}